\documentclass[showpacs,prb,reprint,preprintnumbers,amsmath,amssymb,superscriptaddress,floatfix]{revtex4-1}
\usepackage{graphicx}
\usepackage{dcolumn}
\usepackage{colortbl}
\usepackage{enumitem}

\begin{document}
\newcommand{\oH}{\ensuremath{\mathcal{H}}}
\newcommand{\oI}{\ensuremath{\mathcal{I}}}
\newcommand{\oS}{\ensuremath{\mathcal{S}}}
\newcommand{\oQ}{\ensuremath{\mathcal{Q}}}

\title{Nuclear magnetic resonance inverse spectra of InGaAs quantum dots: Atomistic level structural information}

\author{Ceyhun \surname{Bulutay}}
\email{bulutay@fen.bilkent.edu.tr}
\affiliation{Department of Physics, Bilkent University, Ankara 06800, Turkey}
\author{E. A. \surname{Chekhovich}}
\author{A. I. \surname{Tartakovskii}}
\affiliation{Department of Physics and Astronomy, University of Sheffield, Sheffield S3 7RH, 
United Kingdom}
\date{\today}

\begin{abstract}
A wealth of atomistic information is contained within a self-assembled quantum dot (QD), 
associated with its chemical composition and the growth history. In the presence 
of quadrupolar nuclei, as in InGaAs QDs, much of this is inherited to nuclear spins via 
the coupling between the strain within the polar lattice and the electric quadrupole 
moments of the nuclei. Here, we present a computational study of the recently introduced 
inverse spectra nuclear magnetic resonance technique to assess its suitability for extracting
such structural information. We observe marked spectral differences between the compound InAs and alloy 
InGaAs QDs. These are linked to the local biaxial and shear strains, and the local bonding 
configurations. The cation-alloying plays a crucial role especially for the arsenic nuclei. The 
isotopic line profiles also largely differ among nuclear species: while the central transition of 
the gallium isotopes have a narrow linewidth, those of arsenic and indium are much broader 
and oppositely skewed with respect to each other. The statistical distributions of electric 
field gradient (EFG) parameters of the nuclei within the QD are analyzed. 
The consequences of various EFG axial orientation characteristics are discussed. Finally, 
the possibility of suppressing the first-order quadrupolar shifts is demonstrated by simply 
tilting the sample with respect to the static magnetic field.
\end{abstract}

\pacs{75.75.-c, 76.60.Gv, 76.60.Pc} 

\maketitle

\section{Introduction}
The valuable expertise gained on nuclear spins in quantum dots (QDs) over the past decade has 
revolutionized the traditional research on semiconductor physics.\cite{urbaszek13,chekhovich13,
taylor03,deng05,lai06,hogele12,yang13,economou14,marohn95,gammon97,kikkawa00,makhonin10,
kuznetsova14,munsch13,chekhovich12,makhonin11,dzhioev07,huang10,bulutay12,verbin12,
fallahi10,hogele12,chekhovich14} For instance, solid-state quantum memories, a vital component 
for quantum information technologies, count on the long coherence times of 
nuclear spins.\cite{taylor03} In addition to storing the quantum state,
nuclear spins potentially can act as the central processing unit as in ensemble 
computing\cite{cory97,vandersypen01} that can also be extended to quadrupolar 
nuclei,\cite{yusa05} or, they can present an ideal testbed for quantum control as an integral 
part of an exciton-nuclei feedback loop.\cite{lai06,hogele12,yang13,economou14}

Another emerging utility of QD nuclear spins is for the materials science as a targeted 
nanoscale diagnostic tool. On this front, there have been recent advances in analytical 
techniques such as cross-sectional scanning tunneling microscopy,\cite{bruls02}
coherent x-ray diffraction-based three-dimensional mapping,\cite{kegel00,yacoby02} 
and the atom probe tomography.\cite{kelly07} Routinely, the nuclear magnetic resonance 
(NMR) had been one of the preferred choices in identifying below the parts-per-million-level 
concentration of rare constituents.\cite{levitt-book} This, however works on macroscopic 
samples still of molar sizes. When it comes to probing a \textit{single} QD, the applicability of 
the conventional NMR is hampered because of the insufficient equilibrium magnetizations within 
such small volumes. Added to this, is the low sensitivity inherent in detection by the magnetic 
induction of precessing magnetization.\cite{marohn95} If instead, the already proven optical 
orientation framework is pursued, the electron spin can efficiently polarize the nuclear 
spins within the QD through the contact hyperfine interaction.\cite{paget84} This nuclear 
polarization, known as the Overhauser field, acts back on the exciton and shifts its energy 
as it recombines, leaving a trace on the photoluminescence (PL). Overall, this
makes up the recipe for the NMR of a single QD, which is termed as the optically detected 
NMR (ODNMR).\cite{gammon97} For the detection, it relies on the measurement of either the 
Overhauser field-shifted excitonic PL with a $\mu$eV-resolution,\cite{gammon97} or the Faraday 
rotation in the reflected probe beam with a sensitivity below 1~mrad.\cite{kikkawa00}

It is also desirable to use ODNMR on the widely accessible self-assembled QDs (SAQDs). 
However, an issue that is prevalent in SAQDs is that they inherently possess an inhomogeneous and 
anisotropic strain.\cite{petroff03} In a III-V semiconductor crystal lattice, 
such a strain field causes local electric field gradients (EFG) with which a quadrupolar spin-$I$ 
nucleus, i.e., with $I\ge 1$, interacts because of its electric quadrupole 
moment.\cite{cohen57,das58} This quadrupole interaction (QI) splits the nuclear spin degeneracy 
even in the absence of an external magnetic field, and severely broadens the resonances.
This has posed a challenge for employing standard ODNMR in SAQDs giving rise to poor signal to
noise ratio. Very recently the problem has been alleviated by introducing a 
so-called \textit{inverted} radio frequency (rf) excitation scheme.\cite{chekhovich12} 
This increased substantially the fraction of nuclei participating in the Overhauser shift. 
A breakthrough not only for the atomistic level structural information on strained SAQDs, 
but also for the quantum information technologies. 
This is thanks to the crucial structural information it can supply,
much needed both to engineer a noise-free nuclear spin bath and 
also to perform a coherent control over the Bloch sphere of relatively small number of 
nuclear spins.\cite{makhonin11,chekhovich14} 

After this successful experimental demonstration of the ODNMR inverse spectra on 
strained QDs,\cite{chekhovich12} its full potential awaits to be explored on a 
theoretical level. Therefore, the aim of this work is to undertake a computational 
assessment, choosing InGaAs QDs as the test case. Primarily, we would like to address 
what kind of atomistic level information can possibly be extracted by the technique, 
and where to look for these. 
In particular we give special importance to the central transition lineshape as this is 
experimentally the most conspicuous spectral feature. Moreover, it carries important clues about 
the internal structure of the QD. Thus, we perform a detailed search over 
the parameter space of the inverse spectra technique. This sheds light on the 
line profiles and resolution trade offs, as well as experimentally more impracticable 
aspects such as the dependence of the sample orientation with respect to the magnetic field.
By comparing a binary InAs QD with an identical-shape alloy In$_{0.2}$Ga$_{0.8}$As 
counterpart, we uncover key fingerprints of the alloy composition. Furthermore, we explain 
the basis of these behaviors in terms of available atomistic configurations.
With this insight, we intend to unfold the prospects of the inverse spectra technique 
as a tool to resolve atomistic-level variations in strained nanostructures.

The paper is organized as follows: In Sec.~II we present the theoretical setting of our 
atomistic analysis. In Sec.~III we provide information about our benchmark cases, followed by 
our results. In Sec.~IV we conclude by itemizing our major findings. In the interest of a lucid 
presentation, we defer additional technical supporting materials and figures to a number of appendixes.
Appendix~A discusses nuclear polarizations, and more specifically why it is harder to polarize 
the arsenic nuclei in a strained environment. Appendix~B contains EFG-related histograms for the alloy 
and the compound QDs. In Appendix~C we consider the role of specific quadrupole parameters on the 
lineshape. The last appendix illustrates the effects of individual EFG parameters on spectral 
transitions of a single nucleus to authenticate some of the assertions in the main text.

\section{Theory}

\subsection{Three concomitant coordinate systems}
The crux of our analysis is based on the simultaneous use of a number of coordinate systems.
A QD has a native coordinate system set through the crystallographic axes where
the QD growth axis usually coincides with one of them; in our test cases this is 
the $z$-axis and the [001] direction.
The orientations of an external magnetic field and the optical beam with respect 
to the growth axis of the QD bear particular significance in terms of which
Faraday/Voigt geometries and $\sigma^\pm$ pumping are defined.
Yet, there are
at least two more relevant coordinate axes that gain importance in an atomistic
treatment. Unlike the global crystal axes, these are local, i.e., they change orientation 
with position over the QD.
They are defined through strain and the EFG tensors, denoted
in cubic crystallographic $xyz$ components by $\epsilon_{ij}$ and $V_{ij}$, respectively. 
The two phenomena are linked through $S$, the fourth-rank gradient elastic 
tensor as
\begin{equation}
V_{ij}\equiv\frac{\partial^2V}{\partial x_i\partial x_j}=\sum_{k,l=1}^3 
S_{ijkl}\epsilon_{kl},
\end{equation}
where $V$ is the crystal electric field potential.\cite{shulman57} In the so-called Voigt 
notation, the $S$ tensor for cubic crystals is governed by only two independent components 
$S_{11}$ and $S_{44}$, both of which are experimentally measurable.\cite{sundfors74} 
In cubic crystallographic $xyz$ axes, $S_{11}$ and $S_{44}$ relate the 
diagonal and off-diagonal strain and EFG components, respectively, like
$V_{zz}=S_{11}\epsilon_{B}$, $V_{xy}=2 S_{44} \epsilon_{xy}$, etc. where
$\epsilon_B = \epsilon_{zz}-(\epsilon_{xx}+\epsilon_{yy})/2$ is the so-called
biaxial strain. Similarly, we find it necessary to introduce a shear strain measure as
$\epsilon_S\equiv|\epsilon_{xy}|+|\epsilon_{yz}|+|\epsilon_{zx}|$ 
to quantify the effectiveness of the off-diagonal components.\cite{bulutay12}

The strain and EFG tensors have their own distinct principal axes where each  
becomes diagonal, and within which working with that quantity
becomes highly convenient. Among the three principal axes of a rank-2 quantity (such as 
strain or EFG), the one with the largest absolute value is named as the \textit{major} 
principal axis. Hence, this brings three concomitant coordinate systems at one's disposal.
Our primary interest in nuclear spin states in the presence of QI
favors the explicit use of local EFG principal axes which we shall discriminate by 
the $XYZ$ capital letters,\cite{note-notation} with axes being labeled so as to satisfy the inequalities
$|V_{XX}|\leq|V_{YY}|\leq|V_{ZZ}|$, making $Z$ the major EFG axis.

\subsection{Fundamental Hamiltonian}
In the local $XYZ$ frame the strain-dependent part of the nuclear Hamiltonian responsible
for the QI is given by
\begin{equation}
\oH_Q=\frac{e^2qQ}{4I(2I-1)}\left[ 3\oI^2_Z-\oI^2+\eta\frac{\oI^2_+-\oI^2_-}{2} \right]\, ,
\end{equation}
where $\vec{\mathbf{\oI}}$ is the dimensionless nuclear spin angular momentum vector operator, 
through which we define the above raising/lowering scalar operators 
$\oI_\pm\equiv \oI_X\pm i\oI_Y$. As to the other variables, $Q$ is the electric quadrupole moment
of the nucleus, $q\equiv V_{ZZ}/e$ is the EFG parameter which is also the primary coupling constant 
of QI, with $e>0$ being the electronic charge, and $\eta=(V_{XX}-V_{YY})/V_{ZZ}$ is the biaxiality 
parameter, satisfying $0\le \eta\le 1$ by construction, and it determines the mixing between the 
free nuclear spin magnetic quantum numbers.

In the same local $XYZ$ frame the static magnetic field vector $\mathbf{B}_0$ will be in general 
oblique as described by the spherical polar angles $\theta$, and $\phi$ so that its 
Hamiltonian becomes\cite{das58}
\begin{equation}
\oH_M=-\hbar\Omega\left(\oI_X\sin\theta\cos\phi+\oI_Y\sin\theta\sin\phi+
\oI_Z\cos\theta\right)\, , \nonumber
\end{equation}
where $\Omega\equiv\gamma B_0$, and $\gamma$ is the nuclear gyromagnetic ratio.
Hence, for each nucleus $k$ under consideration, 
$$\left(\oH_Q + \oH_M\right)|i\rangle_k = h \nu^k_i|i\rangle_k \, ,$$
needs to be solved, where we denote the resultant spectrum with $\nu^k_i$, $i=-I,-I+1,\ldots,I$.
Under sufficiently high magnetic fields, which we assume throughout our work, 
dipole-allowed transitions are $i \leftrightarrow i+1$. Among these, the strongest one
$-1/2\leftrightarrow +1/2$ is referred to as the central transition (CT), and the remaining 
weaker ones as the satellite transitions (STs). When the EFG major principal axis deviates from the
$\mathbf{B}_0$ direction, CT becomes broadened only as a second-order effect, hence stays 
quite narrow, whereas STs undergo extensive broadening as they are affected in 
first order.\cite{cohen57}

\subsection{Optical orientation}
We characterize the nuclear spin ensemble within the QD by a nuclear spin temperature 
$T_{\hbox{\tiny nuc}}$ which is a measure of the degree of optical orientation. 
The probability of occupancy of each nuclear spin state, $i$ is governed by a 
thermal distribution of the form
\begin{equation}
p^{\hbox{\tiny th}}_i=\frac{e^{-h\nu^k_i/k_BT_{\hbox{\tiny nuc}}}}
{\sum_{i=-I}^{I}e^{-h\nu^k_i/k_BT_{\hbox{\tiny nuc}}}}\, , \nonumber
\end{equation}
where $k_B$ is the Boltzmann constant.

The helicity of the absorbed optical orientation beam dictates the spin of the 
created electron and hole as a requirement of angular momentum conservation. 
Subsequently, through predominantly the contact hyperfine interaction between 
the electron and nuclear spins, depending on the absorbed photon helicity the 
nuclear spins are either pumped down or up within their individual spectrum $\nu^k_i$, 
toward $i\to -I$ or $i\to +I$, respectively.\cite{urbaszek13} 
In the absence of an rf excitation, and under continuous 
optical orientation this distribution will be sustained.
Therefore, the steady-state is either \textit{normal} for $\sigma^+$, or 
\textit{inverted} for $\sigma^-$ persistent pumping. We represent the former (latter) by using a 
positive (negative) nuclear spin temperature, i.e., $T_{\hbox{\tiny nuc}}>0$ 
($T_{\hbox{\tiny nuc}}<0$).

\subsection{rf Excitation}
The novelty of the inverse spectra technique with respect to conventional saturation spectroscopy
comes from its inverted excitation scheme, which has a white spectrum except for a frequency gap, 
$f_{\hbox{\tiny gap}}$.\cite{chekhovich12} If this gap does not coincide with any of the 
dipole-allowed transitions $\nu^k_i\leftrightarrow\nu^k_{i+1}$, then under a sufficiently long
excitation, the population of all nuclear spin states will be equilibrated at the same value, 
$p^{\hbox{\tiny sat}}_i=1/(2I+1)$, giving rise to zero nuclear polarization, and hence no 
Overhauser shift on the excitonic PL. On the other hand when the gap coincides with one, 
or sometimes simultaneously with a number of transitions, the equilibration will 
only occur within the states that remain under 
the gap-free parts of the excitation. Therefore, the $2I+1$ states will be split into multiple 
groups,\cite{note-multiple} each internally reaching to an individual saturation value
based on the preexisting thermal populations of the member states according to 
\begin{equation}
p^{\hbox{\tiny sat}}_i=\frac{\sum_{i\in G}p^{\hbox{\tiny th}}_i}{N_G}\, ,
\end{equation}
where $G$ is the group index, and $N_G$ is the number of member spin states within that group.

The spin polarization of a nucleus $k$ just after rf excitation will be based on the 
population of each state
\begin{equation}
\vec{P}_k=\frac{1}{I}\sum_{i=-I}^{I} {}_k\langle i| \vec{\mathbf{\oI}} 
|i\rangle_k \, p^{\hbox{\tiny sat}}_i\, . 
\end{equation}
If we denote by $\hat{e}$ the optical pumping direction along which the electronic spin is 
aligned, which is usually, but not always the QD growth axis, then the parallel component of
the nuclear polarization becomes $P^e_k=\hat{e}\cdot\vec{P}_k$. 
In the experiments, the changes in the nuclear polarization are probed via the Overhauser 
energy shift in the excitonic PL signal\cite{urbaszek13} which is given by
\begin{equation}
E_{OS}=\sum_k \left| \psi(\vec{R}_k)\right|^2 A_k I_k P^e_k\, ,
\end{equation}
where $A_k$ is the hyperfine coupling constant, and $\psi(\vec{R}_k)$ is the electronic 
wave function at the nuclear site. Here, we ignore the hole hyperfine interaction
which is about an order of magnitude weaker.\cite{fallahi10}

\section{Results}
\subsection{Test QDs}
\begin{figure}
 \includegraphics[width=0.95\columnwidth]{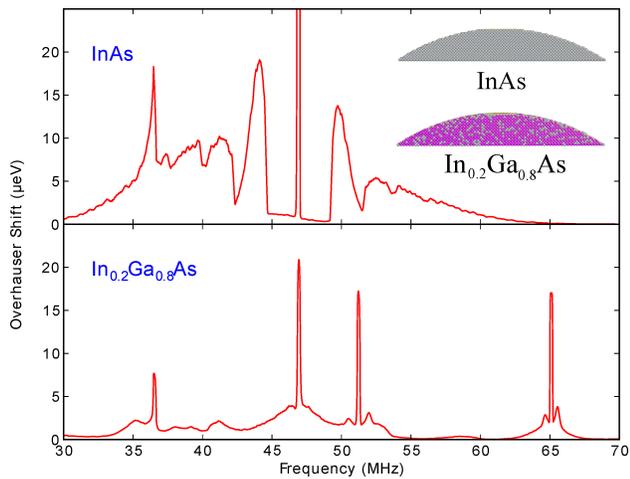}
\caption{(Color online) Inverse NMR spectra of binary InAs, and alloy In$_{0.2}$Ga$_{0.8}$As QDs,
under the conditions $B_0=5$~T, $f_{\hbox{\tiny gap}}$=200~kHz, 
$T_{\hbox{\tiny nuc}}$= 3~mK, $\sigma^+$ optical pumping. Inset shows corresponding QD atoms over 
the (100) cross section.}
\label{binary-vs-alloy}
\end{figure}

The detailed chemical composition profile, i.e., local stoichiometry of InGaAs QDs is still 
an active and  unresolved topic.\cite{biasiol11} A critical factor that nontrivially 
affects the uniformity of the indium distribution within the QD is the annealing process.
It has been reported that the annealed QDs become less uniform along the lateral, but more
uniform along the growth direction; furthermore, the dots get 25\% bigger with respect 
to their pre-annealed sizes in both lateral and growth directions as the indium atoms 
out-diffuse while the gallium atoms diffuse inward.\cite{zibik07}
We base our comparative analysis on two test cases of lens-shaped QDs having a base 
diameter of 40~nm and a height of 6~nm.
Both QDs have an InAs wetting layer and are embedded into a GaAs host 
matrix, but differ in their interior compositions, with one QD being InAs, whereas the 
other being the alloy In$_{0.2}$Ga$_{0.8}$As (see, inset of Fig.~\ref{binary-vs-alloy}),
where indium and gallium atoms are randomly distributed all over the QD region according to
the given mole fraction. Admittedly, these constitute the two extremes, and intermediate cases
like partially segregated alloy realizations are not addressed in this work.
The uniform alloy composition considered here is expected under high growth rate
conditions, where the landing atoms on the surface do not have time to segregate into binary 
compounds, as they quickly get covered by the next layer.\cite{vastola11} 
The rationale behind the selection of these two cases is based on their distinct strain, 
and hence quadrupolar characteristics.\cite{bulutay12}

The computational supercell contains more than 2 million atoms, most of them 
residing in the host matrix, and the QD itself has 171,884 atoms.
We follow the procedure presented in Ref.~\onlinecite{bulutay12} for the relaxation of the 
embedded QDs to their final structures, and the extraction of the atomistic strain distributions.
The only exception in the present work is that we do not perform any nearest-neighbor strain averaging
as this would hinder the true linewidths of the isotope-dependent NMR spectra. 

\subsection{General spectral aspects}
The inverse spectra for both test cases are shown in Fig.~\ref{binary-vs-alloy} computed with the 
associated parameters of $B_0=5$~T, $f_{\hbox{\tiny gap}}$=200~kHz, and $T_{\hbox{\tiny nuc}}$= 3~mK,
chosen to be representative of a realistic case.\cite{chekhovich12} The compound and alloy QD inverse
spectra are strikingly different. The spectra in Fig.~\ref{binary-vs-alloy} is 
the cumulative result of all nuclei within the QD, under $\sigma^+$ optical pumping. 
In all cases, unless stated otherwise, the static magnetic field vector, QD growth axis, 
and the optical pump beam directions are all collinear. Note that as we would like to develop a 
basic understanding of a typical InGaAs QD inverse spectra, throughout this work we use 
a simple \textit{uniformly distributed} electron wave function that is confined within the 
lens-shaped QD region. 

\begin{figure}
 \includegraphics[width=\columnwidth]{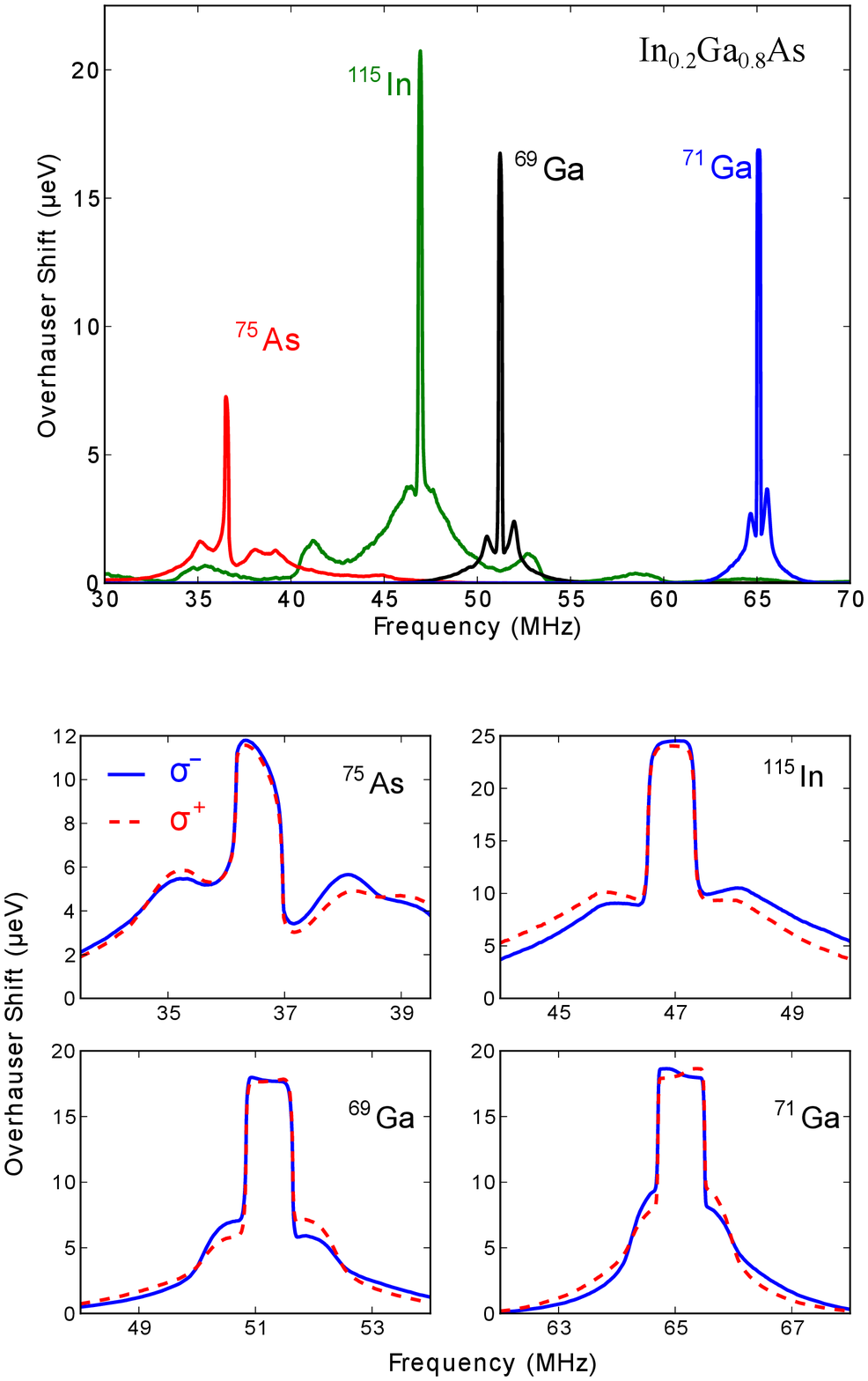}
\caption{(Color online) (Upper panel) The contribution of individual isotopic species
under the same conditions as of Fig.~\ref{binary-vs-alloy}.
(Lower panel) The effect of optical pumping helicity ($\sigma^+$/$\sigma^-$) on the inverse spectra, 
under the same conditions as Fig.~\ref{binary-vs-alloy} other than $f_{\hbox{\tiny gap}}$=800~kHz.}
\label{isotope-helicity}
\end{figure}

\begin{figure*}
 \includegraphics[width=0.84\paperwidth]{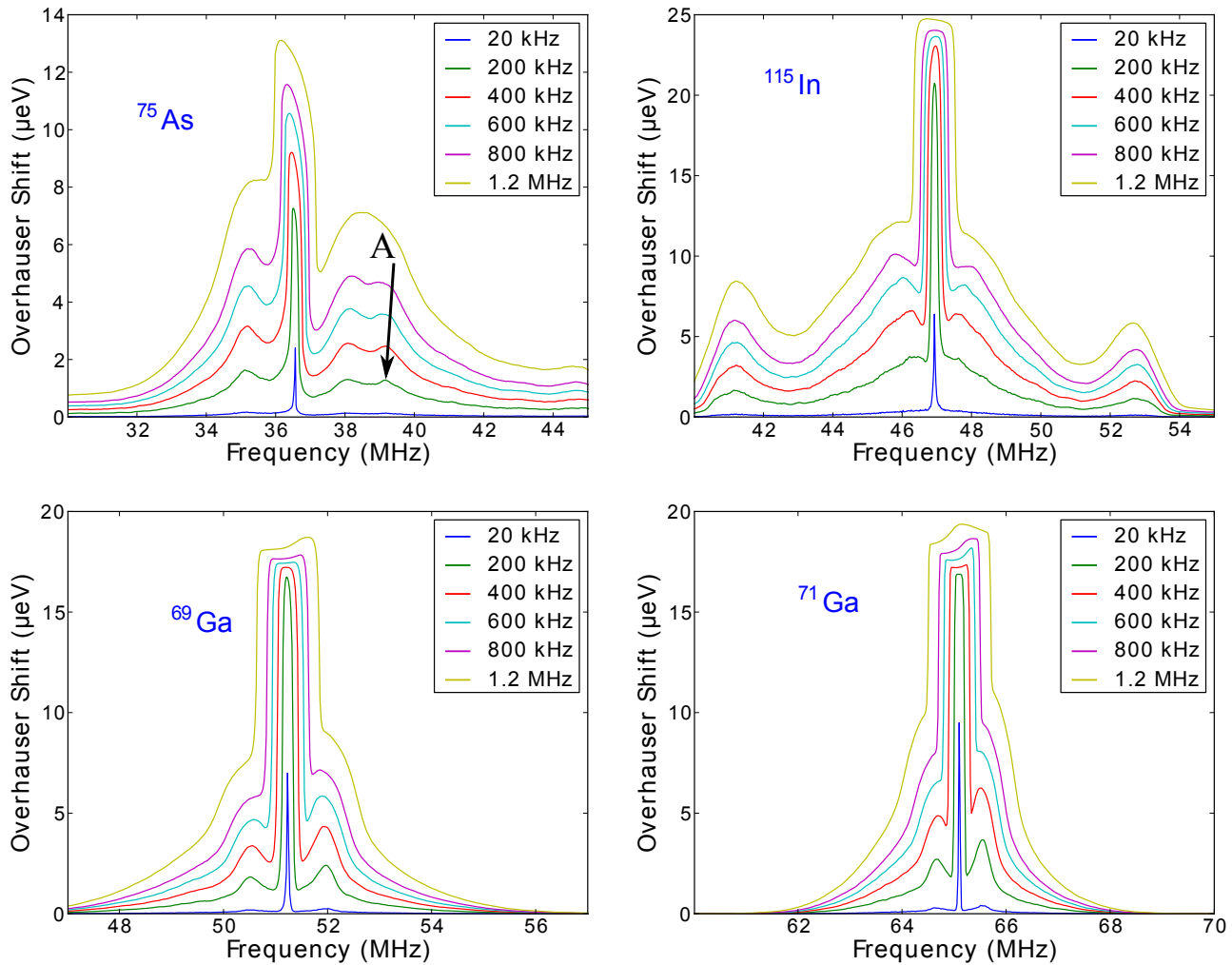}
\caption{(Color online) The evolution of alloy In$_{0.2}$Ga$_{0.8}$As QD 
inverse spectra with respect to $f_{\hbox{\tiny gap}}$ for $\sigma^+$ optical pumping, 
$B_0=5$~T, $T_{\hbox{\tiny nuc}}$= 3~mK.}
\label{fgap-scan}
\end{figure*}

The contribution of individual isotopic species 
as well as the dependence on the light helicity with respect to external magnetic field
are displayed in Fig.~\ref{isotope-helicity}. 
Even though the number of arsenic nuclei is the largest, their CT resonance has the 
smallest peak, the reason for which is related to the hardness in polarizing the arsenic
nuclei as elaborated further in Appendix~A. The indium nuclei because of their 9/2 spins, 
have much extended STs as can be observed from the upper panel of 
Fig.~\ref{isotope-helicity}. The spectrum asymmetry of the neighboring STs on either 
side of the CT are seen to be switched by changing pumping helicity (see, Fig.~\ref{isotope-helicity} 
lower panel), as a matter of fact the contrast can be enhanced further by increasing the initial 
polarization which amounts to lowering of $T_{\hbox{\tiny nuc}}$. Also note that to boost the small 
differences, here we prefer to use a larger $f_{\hbox{\tiny gap}}$ value of 800~kHz. 

In experiments, the choice of $f_{\hbox{\tiny gap}}$ value can indeed become a crucial 
decision for the inverse spectra. To highlight the trade off between spectral resolution 
and the signal intensity, in Fig.~\ref{fgap-scan} we display the spectral evolution as 
a function of $f_{\hbox{\tiny gap}}$ for each isotopic species. The resolution-limited 
flat-top profiles quickly emerge for the gallium nuclei indicating their 
narrow linewidths as will be analyzed below in more depth. On the other hand, for indium and 
especially arsenic nuclei, a large $f_{\hbox{\tiny gap}}$ value may still be preferred 
which is particularly beneficial to capture the relatively weak features associated with 
the STs. One example for this is the emergence under larger $f_{\hbox{\tiny gap}}$ values 
of an additional ST peak as indicated by an arrow on the top left arsenic panel of 
Fig.~\ref{fgap-scan}. For a spin-3/2 system only three peaks are expected, namely,
3/2$\to$1/2, 1/2$\to$-1/2, and -1/2$\to$-3/2. Therefore, this fourth peak which 
unambiguously belongs to arsenic nuclei (see, Fig.~\ref{isotope-helicity}, top panel) 
is rather curious. We identify it as the \textit{alloy peak} with a reasoning based on 
an atomistic configuration analysis, however deferring its detailed discussion for now.

\begin{figure*}
 \includegraphics[width=0.84\paperwidth]{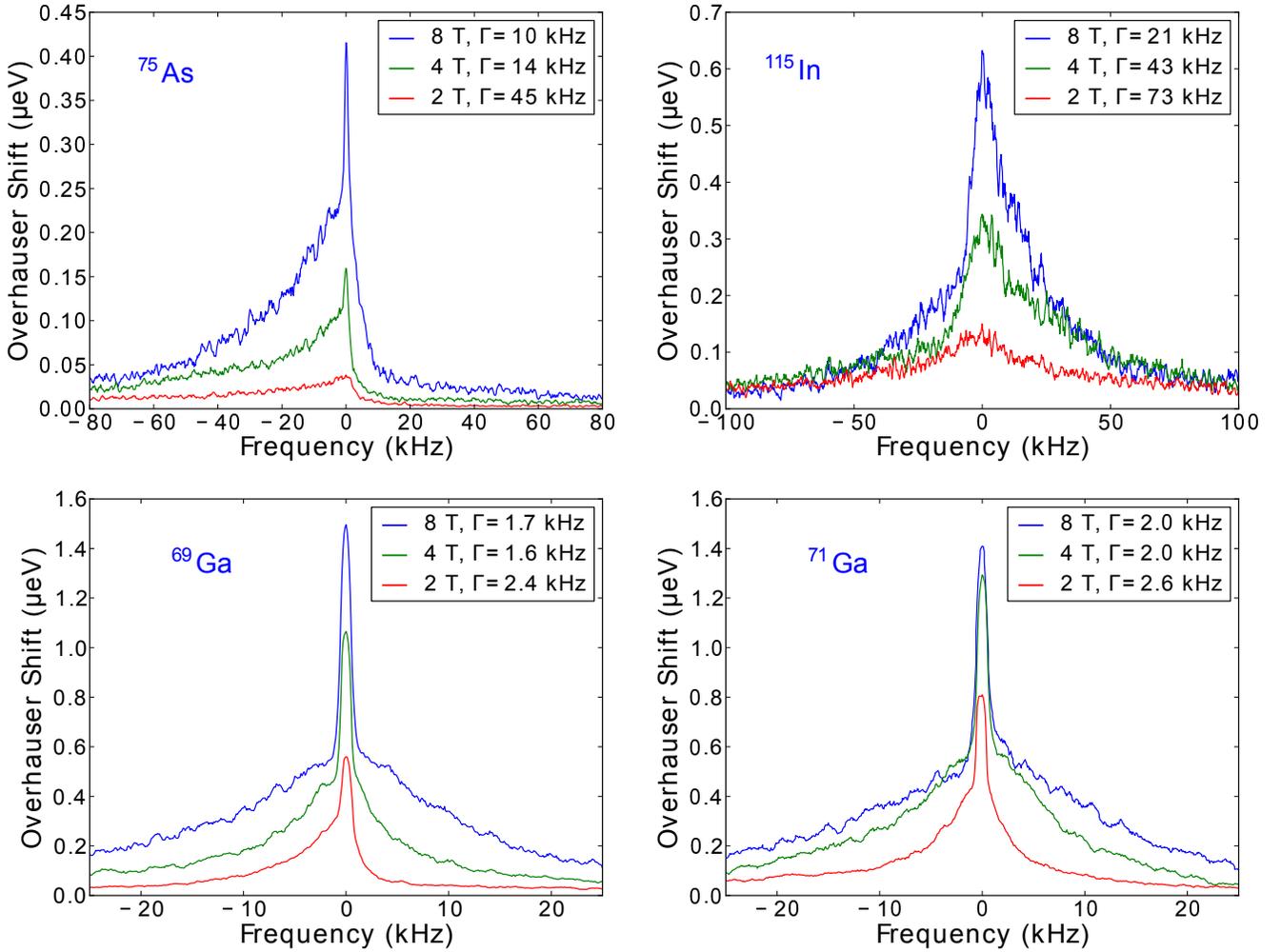}
\caption{(Color online) The dependence of CT line profiles on the external magnetic field for 
all the isotopes. Full width at half maximum ($\Gamma$) values are included in the legend boxes.
To superimpose their peaks, curves for different magnetic fields are displaced in frequency. 
Alloy In$_{0.2}$Ga$_{0.8}$As QD is considered with $f_{\hbox{\tiny gap}}$=1~kHz.}
\label{CT-isotopes}
\end{figure*}

\subsection{Central transition linewidth and profile}
The CT lineshape is one of the means to probe information on the nuclear spin environment. 
For this purpose, first we select a small $f_{\hbox{\tiny gap}}=1$~kHz,
which is ultimately limited by the nuclear homogeneous linewidth.\cite{homogeneous}
Considering alloy In$_{0.2}$Ga$_{0.8}$As QD, CT line profiles at different magnetic
fields are shown in Fig.~\ref{CT-isotopes}. We should note that 
as there always remains some residual overlap from the ST of the other isotopic species,
here the individual \textit{isotopic} contributions, and not the \textit{total} signals are plotted. 
These isotopic line profiles display distinct features, namely, 
$^{69}$Ga and $^{71}$Ga both have quite narrow main peaks over a broad pedestal,
while In and As have evidently opposite \textit{asymmetric} lineshapes.

\begin{figure}
 \includegraphics[width=1.1\columnwidth]{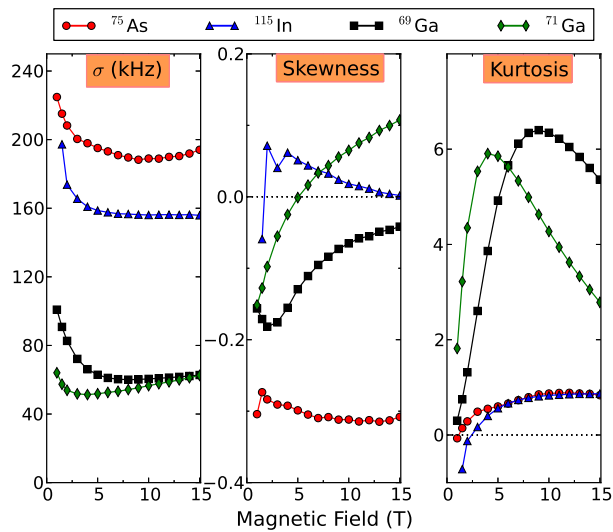}
\caption{(Color online) Standard deviation ($\sigma$), 
skewness, and kurtosis of the CT for all isotopes contained in the alloy 
In$_{0.2}$Ga$_{0.8}$As QD. The inverse spectra are computed with $f_{\hbox{\tiny gap}}=1$~kHz.}
\label{moments}
\end{figure}

To quantify these trends, we make use of the mean, variance, skewness and 
kurtosis\cite{NR} of these distributions; for a sample $\{x_j\}$ of $N$ data points  
these quantities are, respectively
\begin{eqnarray*}
\bar{x} & = & \frac{1}{N}\sum_j x_j \, , \\
\mathrm{Var}(x) & = & \frac{1}{N-1}\sum_j \left( x_j-\bar{x}\right)^2 \, , \\
S_{kp1} & = & \frac{\bar{x}-M}{\sigma} \, , \\
\mathrm{Kurt}(x) & = & \left[ \frac{1}{N}\sum_j \left( 
\frac{x_j-\bar{x}}{\sigma}\right)^4\right]-3 \, ,
\end{eqnarray*}
where among a number of alternatives we prefer Pearson's\cite{pearson1895} 
mode skewness coefficient, $S_{kp1}$, with $\sigma=\sqrt{\mathrm{Var}(x)}$ 
being the standard deviation, and $M$ is the mode (peak value) of the distribution.
Skewness is a dimensionless asymmetry parameter; for unimodal cases zero skewness 
corresponds to a symmetric
distribution around its mode so that tails on either side balance out. Kurtosis is
a dimensionless measure of relative peakedness or flatness of a distribution; for
a normal distribution it becomes zero.

We employ these shape quantifiers on the CT of alloy In$_{0.2}$Ga$_{0.8}$As QD for all 
isotopes as illustrated in Fig.~\ref{moments} which not only corroborate well with the
observations of Fig.~\ref{CT-isotopes} but also reveal some additional trends. 
For all isotopes the standard deviations diverge as the magnetic field decrease below 2~T
into the QI-dominant regime. The $^{71}$Ga isotope has the narrowest $\sigma$
which is also matched by $^{69}$Ga at high magnetic fields.
Regarding skewness, $^{71}$Ga monotonically changes its asymmetry from red- to 
blue-tailed making a transition at 5~T. As was qualitatively noted from Fig.~\ref{CT-isotopes}, 
it is quantitatively asserted in Fig.~\ref{moments} that In and As possess opposite skewness, 
a point which we shall discuss further after we lay out the atomistic structural analysis. Finally, from
the kurtosis panel we observe that all isotopes start from a flat distribution 
within the QI-dominant regime at low magnetic fields which evolves to a peaked 
shape at higher fields. The gallium isotopes go through maxima around 4~T and 9~T 
for $^{71}$Ga and $^{69}$Ga, respectively.

\subsection{Alloy bonding and consequences of quadrupole axial tilting}
As illustrated in Appendix~B through a number of histograms, the spread of quadrupole 
axial tilting is most pronounced for the As nuclei among all elements.
This can be linked to the combined effect of the large \textit{variation} 
in the shear strain component, $\epsilon_S$ of As atoms, compounded by the particularly 
high $S_{44}$ value of As nuclei that is more than 2.5 times of those of Ga and In values;
see, also Appendix~C.
The large variance in $\epsilon_S$ of As has a chemical origin which stems from the mixed 
cation neighbors in the tetrahedral bonding: center As atom is coordinated with a different 
number of Ga/In atoms depending on the local alloy realization. The corresponding tilt angles, 
$\theta$ of the major EFG axes in each bonding configuration are presented in Table~\ref{arsenic-configs}.
This is in support of a recent NMR study which concludes that $^{75}$As QI is highly sensitive 
to different cation coordinations.\cite{knijn10} In the case of In or Ga atoms their 
nearest neighbors are always As, thus, the local strain variation in the cations 
(Ga, In) is more of a \textit{next-nearest-neighbor} effect. 

\textit{What is the physical implication of large variance in quadrupole axial tilting?}
The nuclear dipole-dipole interaction is the main channel for nuclear spin diffusion via 
pairwise flip-flops. However, if the major quadrupolar alignment of each of the involved 
nuclei is significantly off, this inhibits a flip-flop event on the basis of 
energy mismatch. We noted above that the alloy QD and in particular the As nuclei
have a much wider variation in quadrupole axial tilting due to change in the local neighborhood 
as compared to the compound QD  case. Therefore, QDs with large variance in shear strain are 
ideal candidates for reduced nuclear spin diffusion, hence prolonged $T_2$ times, as 
validated by recent experiments (see, Ref.~\onlinecite{chekhovich14}, and references therein).
In fact, in resemblance to \textit{defect centers} that receive wide attention for 
spintronics applications,\cite{koenraad11,doherty13} a random alloy In$_{x}$Ga$_{1-x}$As 
QD, especially of low molar fraction can be termed as a \textit{defect colony} with so many indium 
atoms replacing the gallium of the host lattice. 

\begin{table}
\caption{(Color online) Major quadrupole axis orientations, denoted by the angle $\theta$
with respect to the static magnetic field (in these figures, along the vertical direction),
for all possible arsenic-centric configurations at their pre-relaxation stages. 
Note that consecutive local strain relaxation (cf. Fig.~\ref{histogram} in Appendix~B)
will result in a variance around these $\theta$ values.}
\label{arsenic-configs}
\begin{center}  
\begin{tabular}{|c|c|c|c|}
\hline
 \begin{tabular}{c}
  \footnotesize{$\#1$}\\
  \includegraphics[scale=0.08]{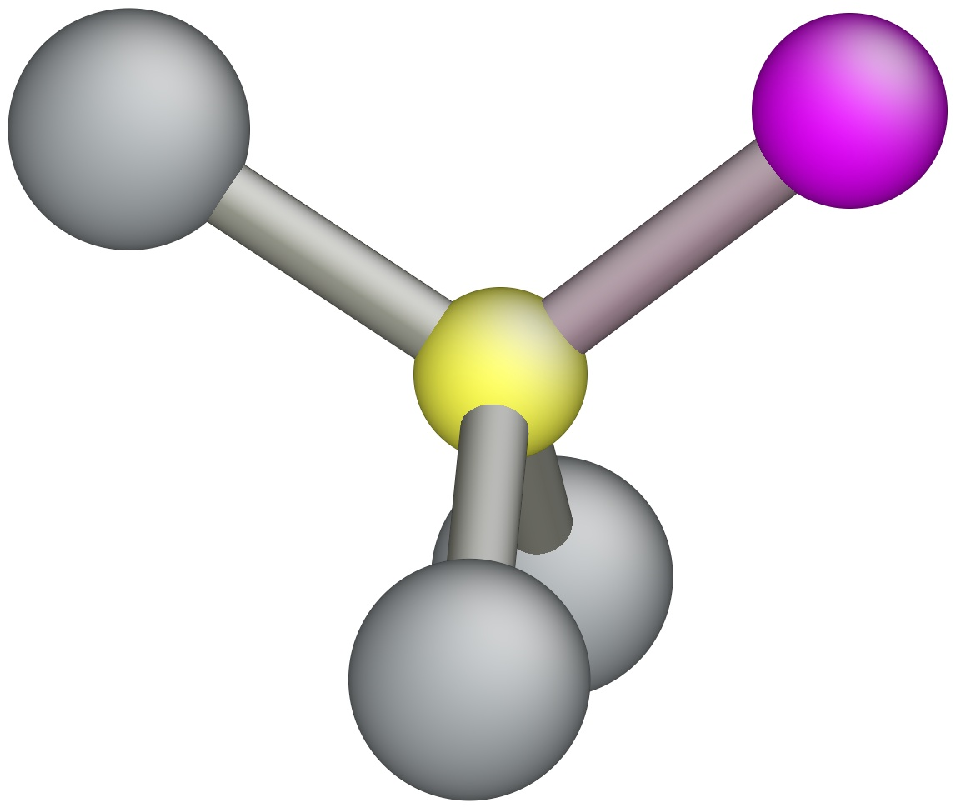}\\
  \footnotesize{$\theta=54.7^\circ$}
 \end{tabular}
&
 \begin{tabular}{c}
  \footnotesize{$\#2$}\\
  \includegraphics[scale=0.08]{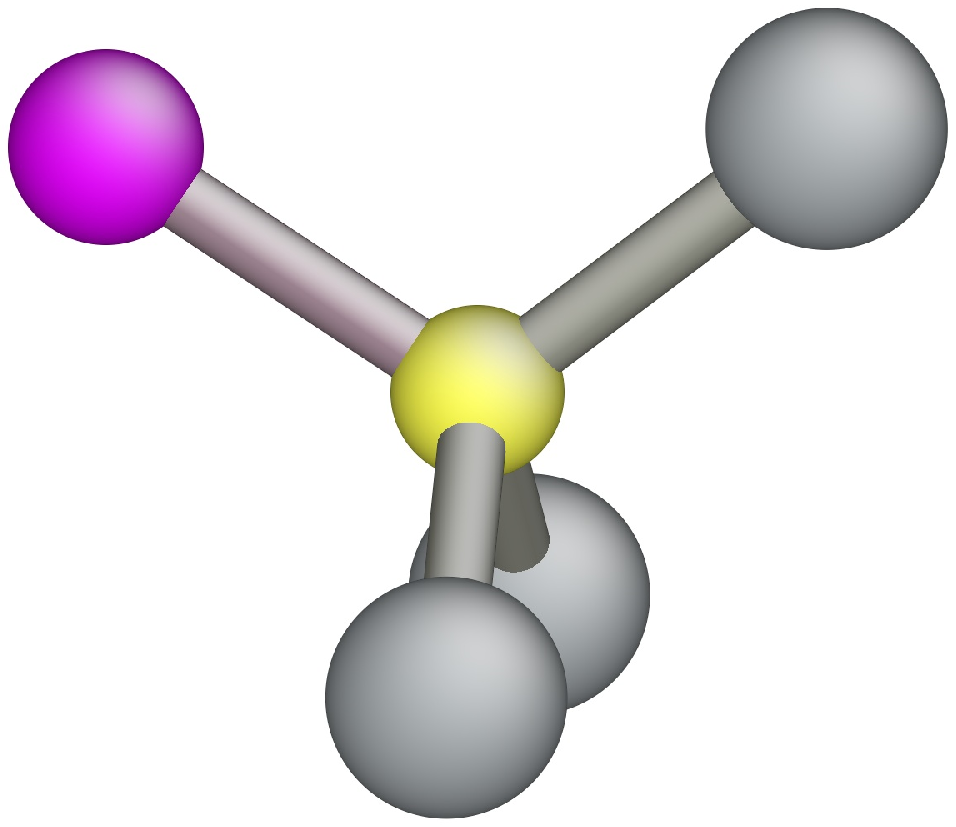}\\
  \footnotesize{$\theta=54.7^\circ$}
 \end{tabular}
&
 \begin{tabular}{c}
  \footnotesize{$\#3$}\\
  \includegraphics[scale=0.08]{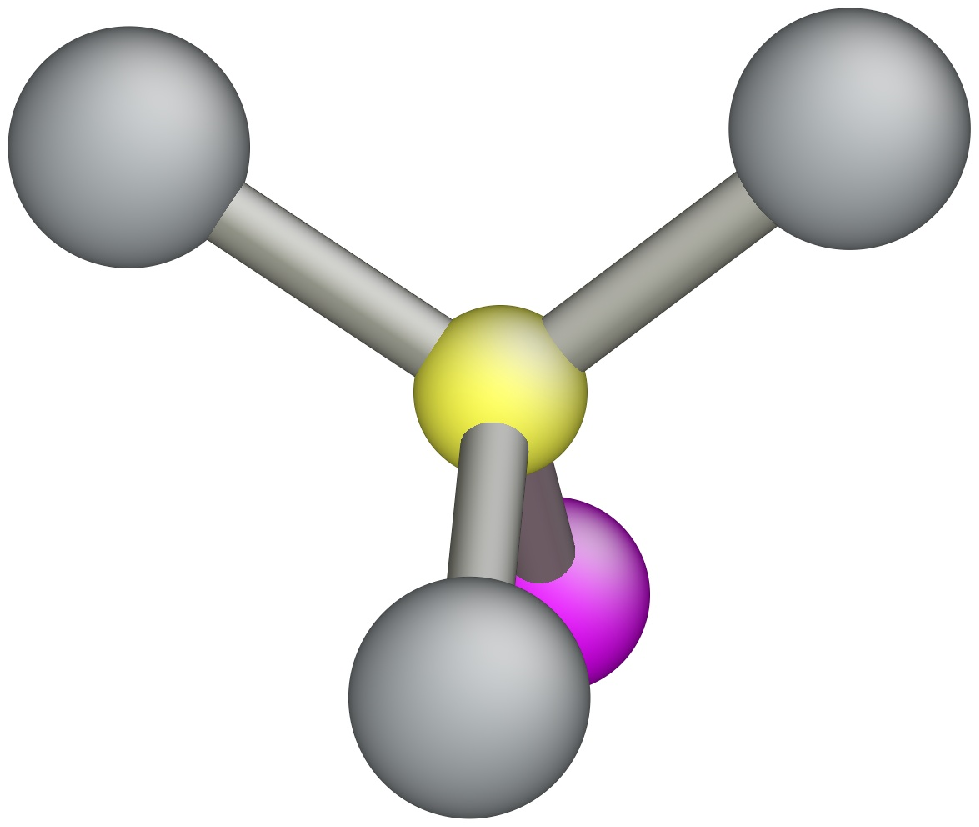}\\
  \footnotesize{$\theta=54.7^\circ$}
 \end{tabular}
&
 \begin{tabular}{c}
  \footnotesize{$\#4$}\\
  \includegraphics[scale=0.08]{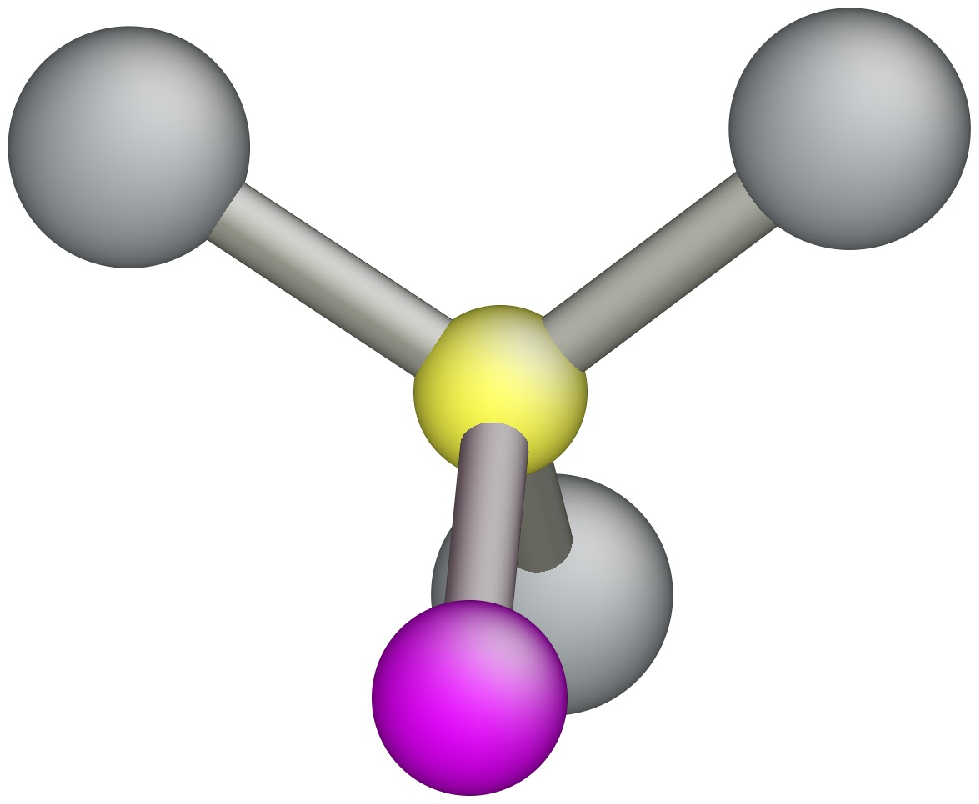}\\
  \footnotesize{$\theta=54.7^\circ$}
 \end{tabular}
\\
\hline
 \begin{tabular}{c}
  \footnotesize{$\#5$}\\
  \includegraphics[scale=0.08]{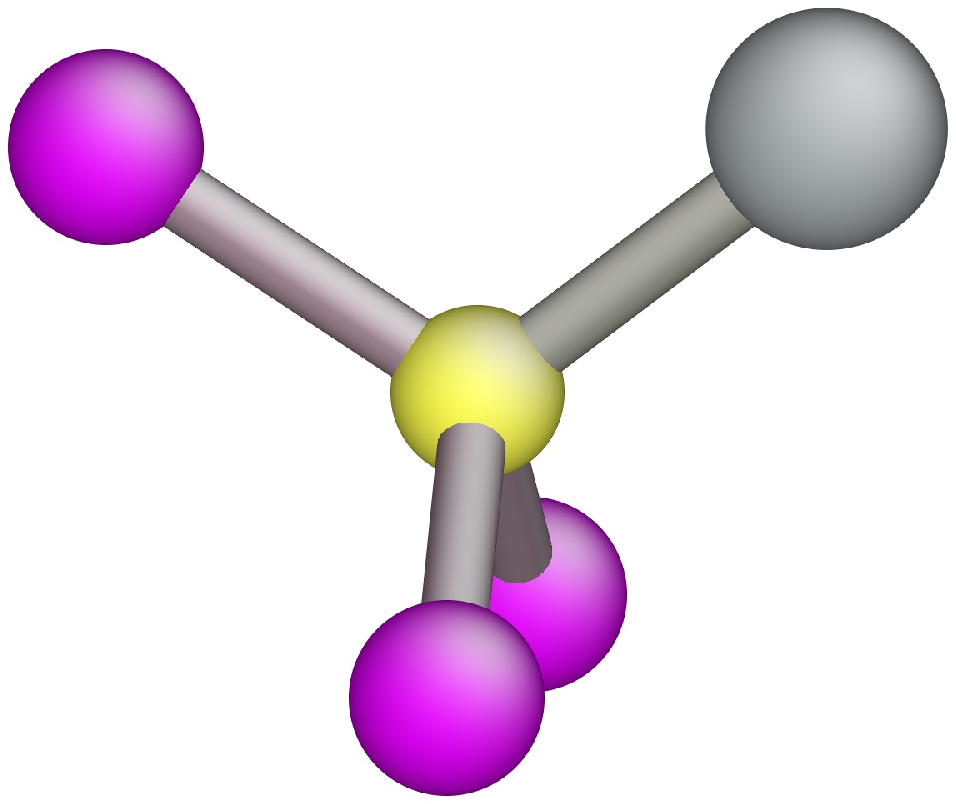}\\
  \footnotesize{$\theta=54.7^\circ$}
 \end{tabular}
&
 \begin{tabular}{c}
  \footnotesize{$\#6$}\\
  \includegraphics[scale=0.08]{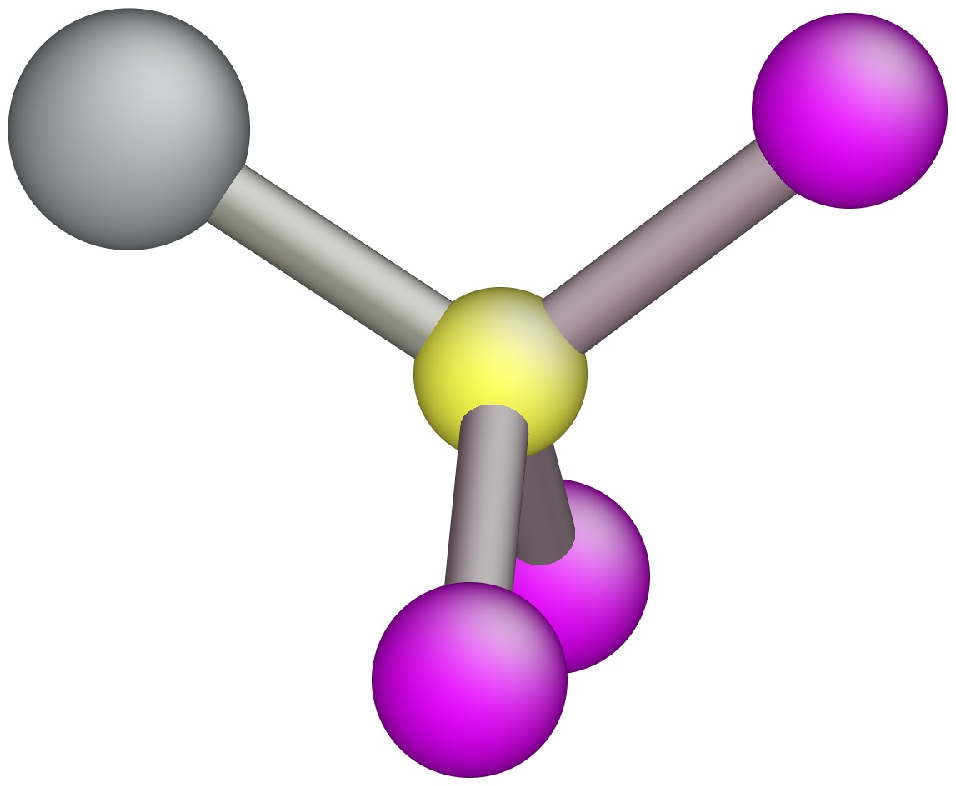}\\
  \footnotesize{$\theta=54.7^\circ$}
 \end{tabular}
&
 \begin{tabular}{c}
  \footnotesize{$\#7$}\\
  \includegraphics[scale=0.08]{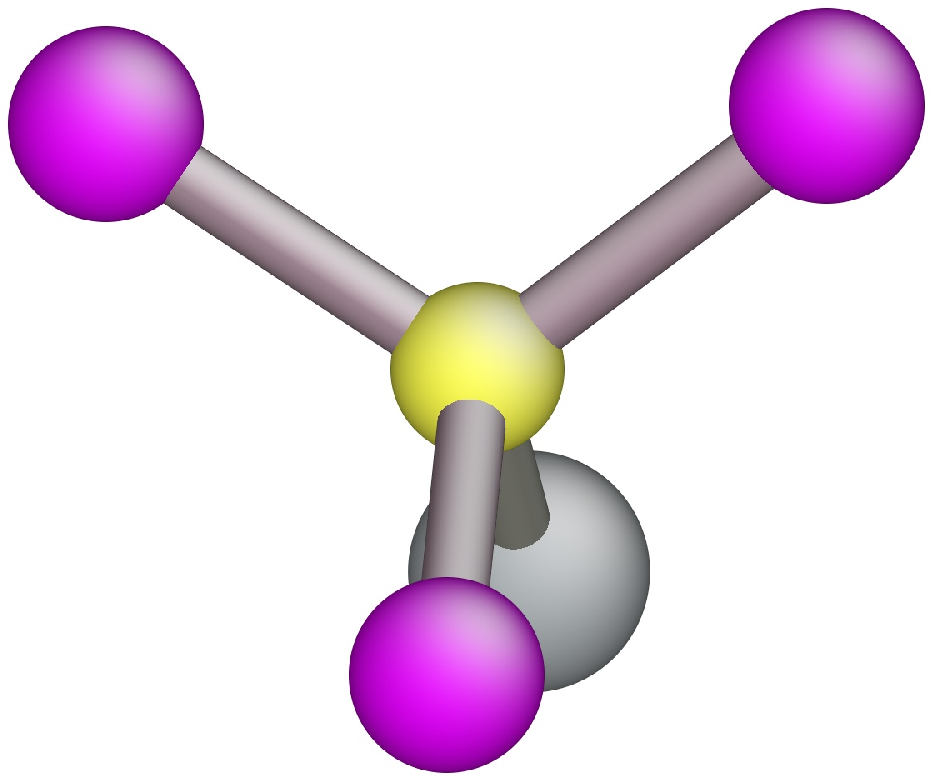}\\
  \footnotesize{$\theta=54.7^\circ$}
 \end{tabular}
&
 \begin{tabular}{c}
  \footnotesize{$\#8$}\\
  \includegraphics[scale=0.08]{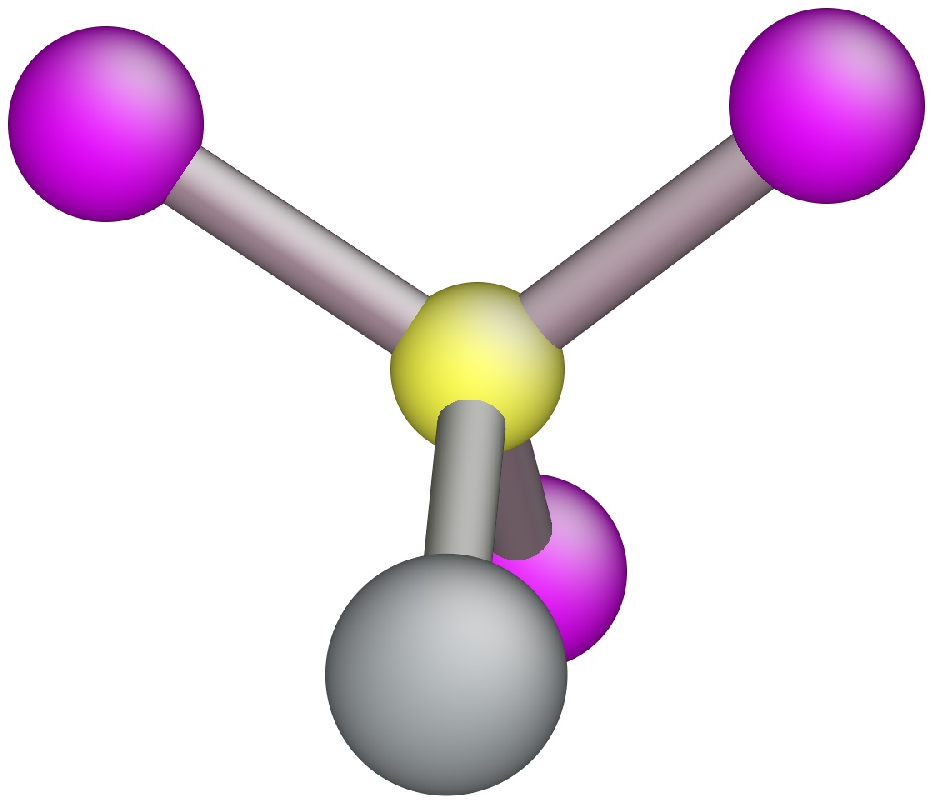}\\
  \footnotesize{$\theta=54.7^\circ$}
 \end{tabular}
\\
\hline
 \begin{tabular}{c}
  \footnotesize{$\#9$}\\
  \includegraphics[scale=0.08]{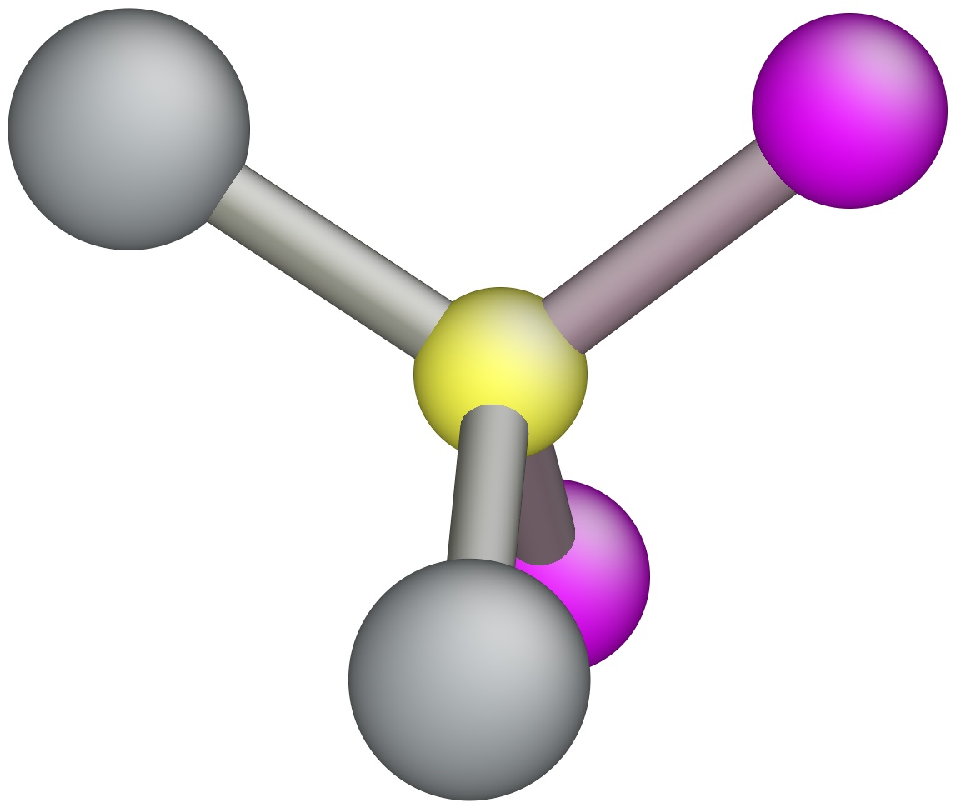}\\
  \footnotesize{$\theta=45.0^\circ$}
 \end{tabular}
&
 \begin{tabular}{c}
  \footnotesize{$\#10$}\\
  \includegraphics[scale=0.08]{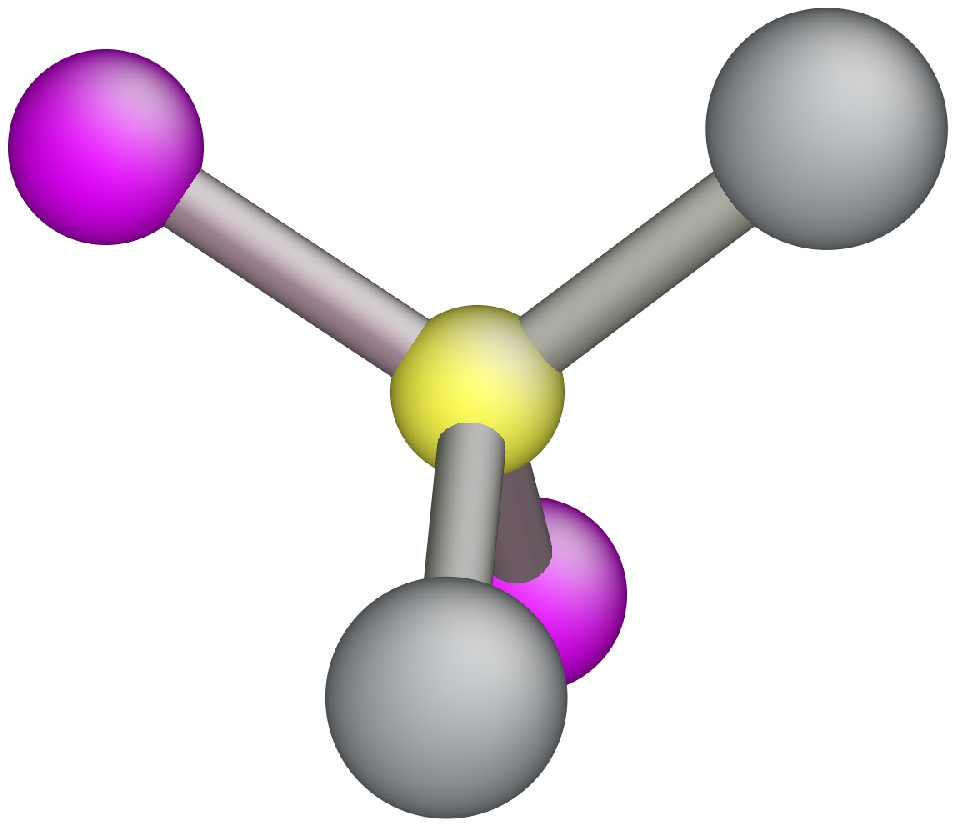}\\
  \footnotesize{$\theta=45.0^\circ$}
 \end{tabular}
&
 \begin{tabular}{c}
  \footnotesize{$\#11$}\\
  \includegraphics[scale=0.08]{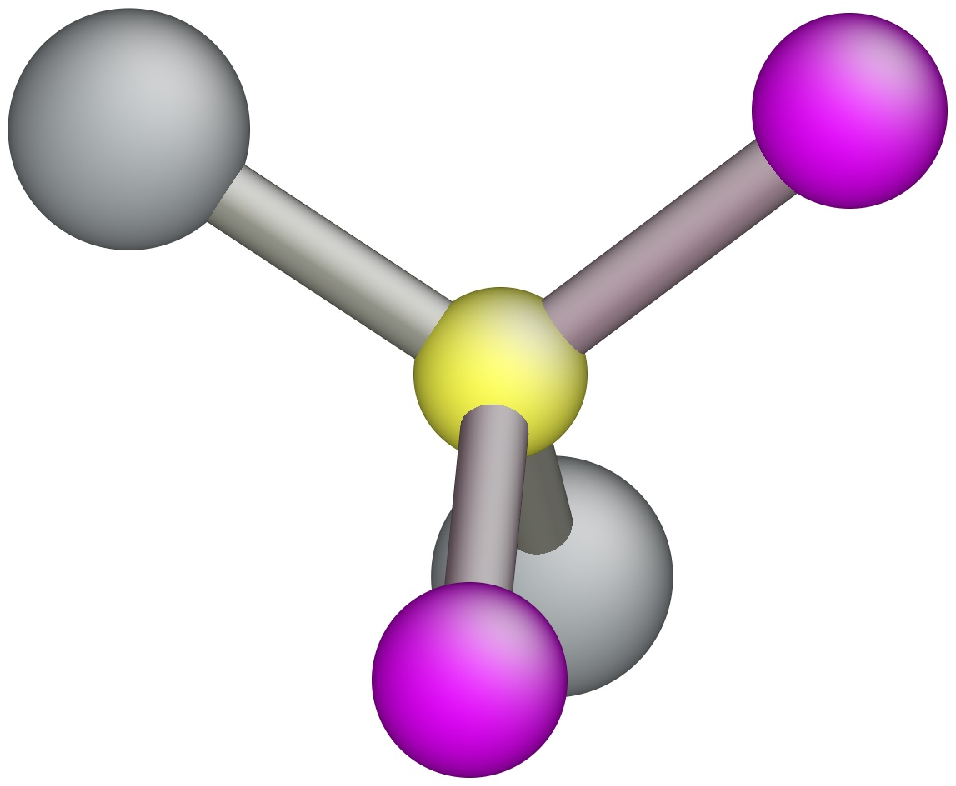}\\
  \footnotesize{$\theta=45.0^\circ$}
 \end{tabular}
&
 \begin{tabular}{c}
  \footnotesize{$\#12$}\\
  \includegraphics[scale=0.08]{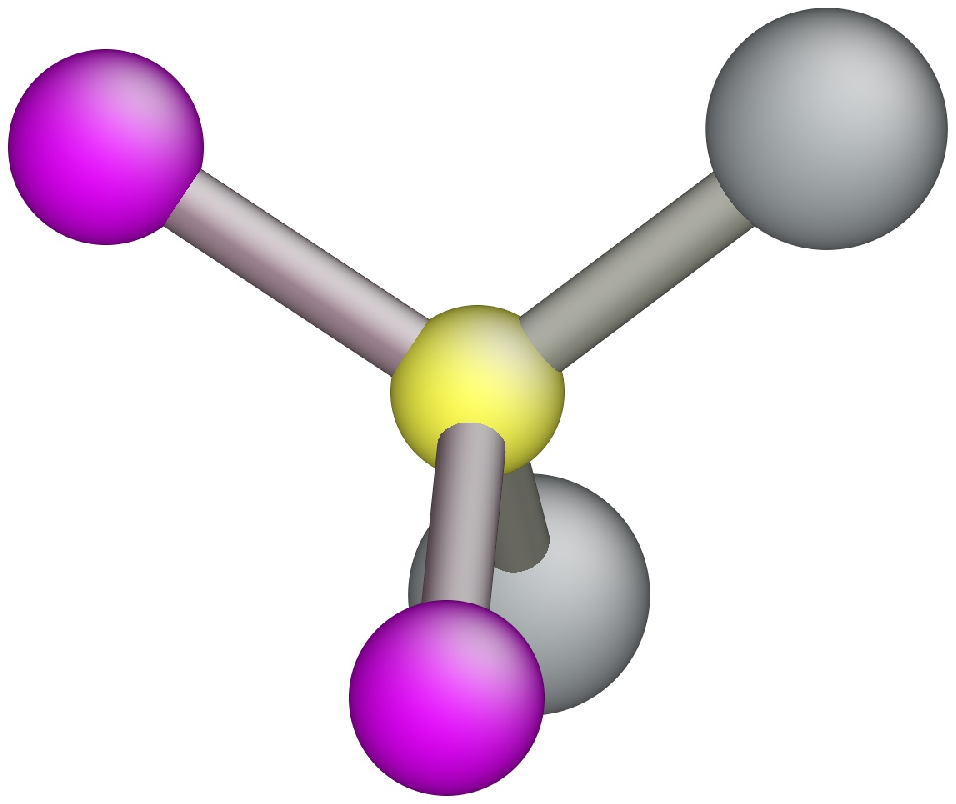}\\
  \footnotesize{$\theta=45.0^\circ$}
 \end{tabular}
\\
\hline
 \begin{tabular}{c}
  \footnotesize{$\#13$}\\
  \includegraphics[scale=0.08]{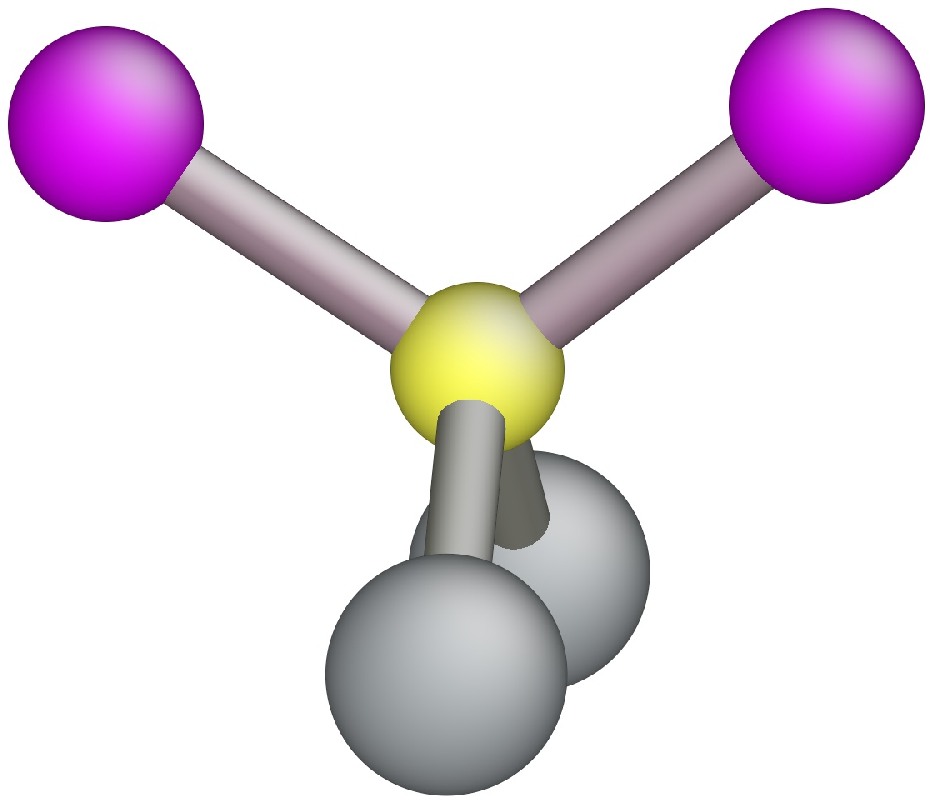}\\
  \footnotesize{$\theta=89.6^\circ$}
 \end{tabular}
&
 \begin{tabular}{c}
  \footnotesize{$\#14$}\\
  \includegraphics[scale=0.08]{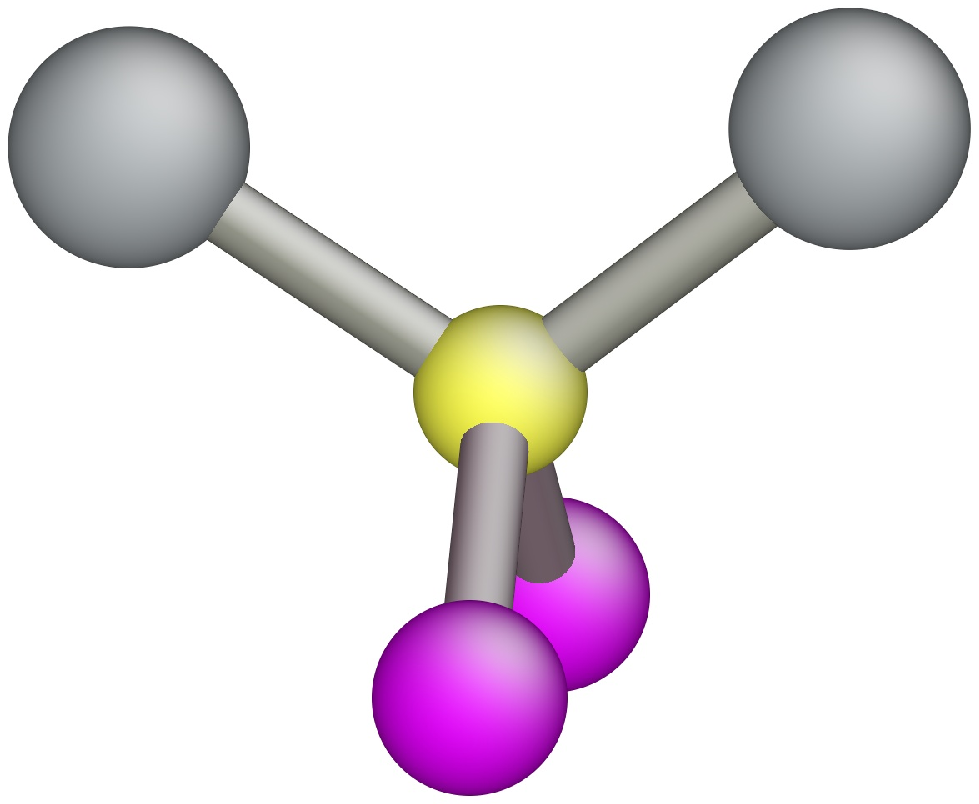}\\
  \footnotesize{$\theta=89.6^\circ$}
 \end{tabular}
&
 \begin{tabular}{c}
  \footnotesize{$\#15$}\\
  \includegraphics[scale=0.08]{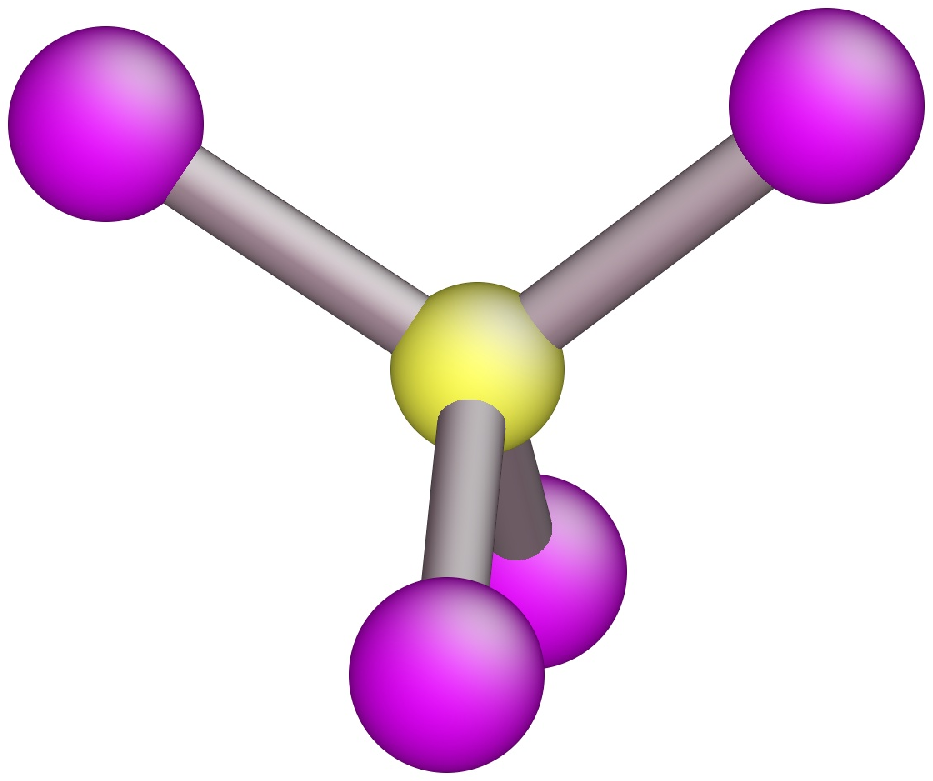}\\
  \footnotesize{$\theta=$N/A}
 \end{tabular}
&
 \begin{tabular}{c}
  \footnotesize{$\#16$}\\
  \includegraphics[scale=0.08]{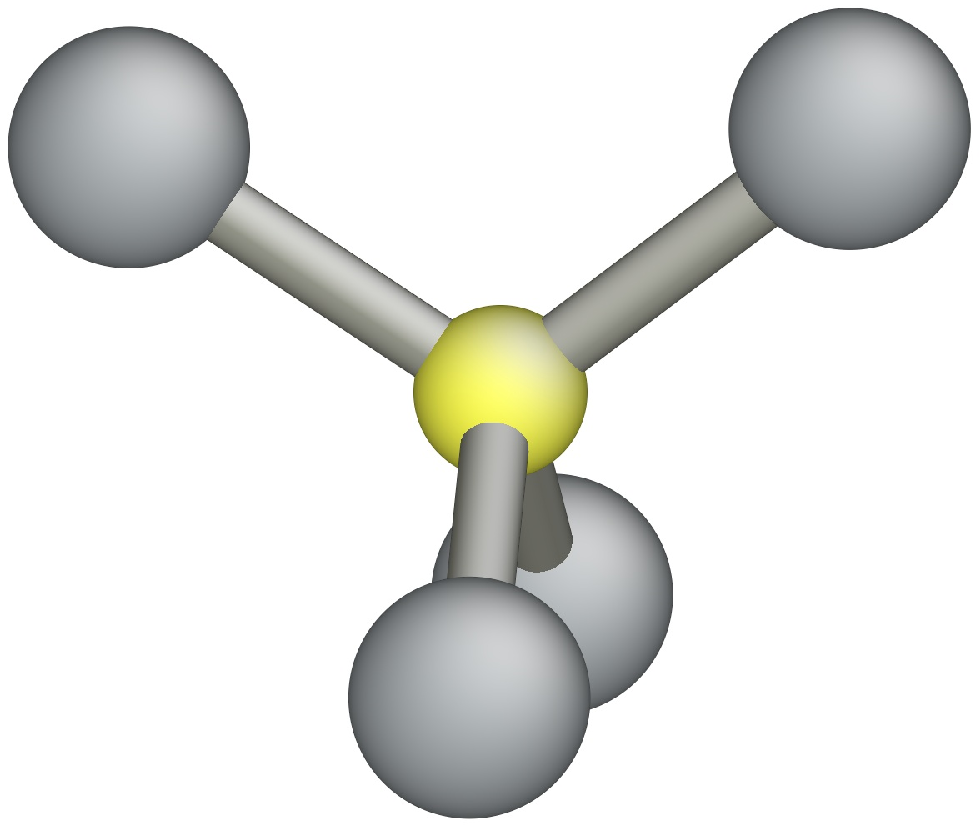}\\
  \footnotesize{$\theta=$N/A}
 \end{tabular}
\\
\hline
\end{tabular}  
\end{center}  
\end{table} 

\begin{figure*}
 \includegraphics[width=0.84\paperwidth]{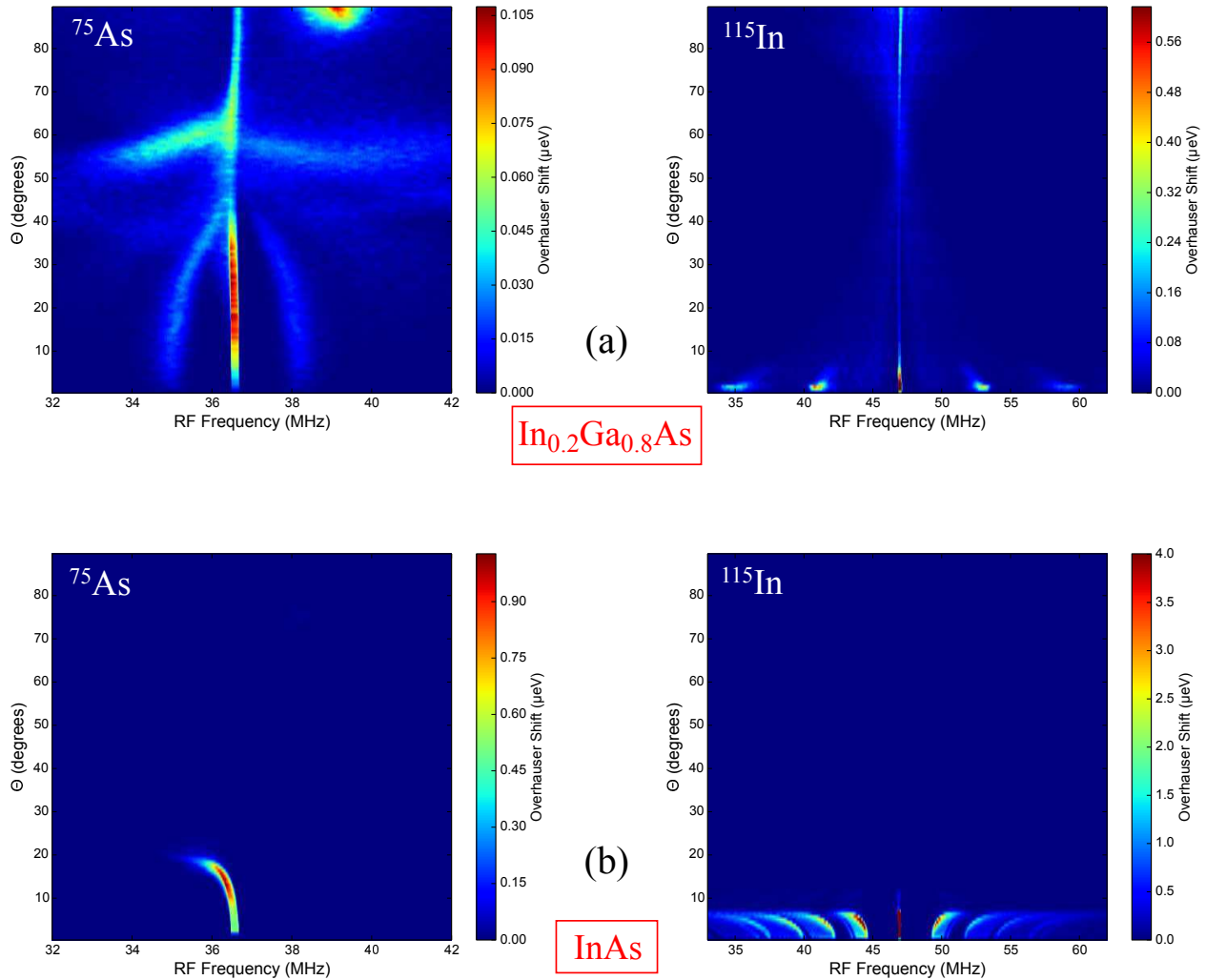}
\caption{(Color online) Contributions to inverse spectra from the nuclei as a function of their
major EFG orientations with respect to static magnetic field, denoted by the angle $\theta$. 
(Left panels) $^{75}$As nuclei; (right panels) $^{115}$In nuclei.
(a) Alloy In$_{0.2}$Ga$_{0.8}$As QD, (b) InAs QD.
For all cases $B_0=5$~T, $f_{\hbox{\tiny gap}}$=200~kHz, and $T_{\hbox{\tiny nuc}}$= 3~mK
under $\sigma^+$ optical pumping. For the panels on the right, to improve visibility the
color scale maxima are set to a quarter of their ordinary values.
}
\label{EFG-2d}
\end{figure*}

\subsection{Hallmark for random alloying in ST band}
We now return to the additional ST peak on the top left $^{75}$As panel of 
Fig.~\ref{fgap-scan} (marked with A and an arrow) which is unexpected for a spin-3/2 system. 
We attribute this A-peak to the cation-alloying present within the QD. Specifically we trace
its origin to the As nuclei with their major quadrupolar EFG axes tilted \textit{perpendicular} to the
magnetic field. This can be observed from Fig.~\ref{EFG-2d} where we analyze the contribution of 
nuclei tagged with respect to their EFG axial tilting. In the case of In nuclei (right panels), 
we do not see a particularly distinct feature coming from the nuclei (very few in number) that 
are close to perpendicular orientation ($\theta=90^\circ$). In contrast, those for the case of As 
nuclei within the alloy In$_{0.2}$Ga$_{0.8}$As QD (left top panel) are clearly responsible for the A-peak. 
Because of the rather different strain environment of these nuclei, a distinct peak occurs (around 39~MHz)
markedly separated in frequency from the neighboring ST peak (around 38~MHz). For the InAs QD, as there 
is only cation-alloying on the interfaces, there are almost no perpendicularly-tilted As nuclei 
(bottom left panel in Fig.~\ref{EFG-2d}), hence no contribution from them or an A-peak.

\begin{figure*}
 \includegraphics[width=0.84\paperwidth]{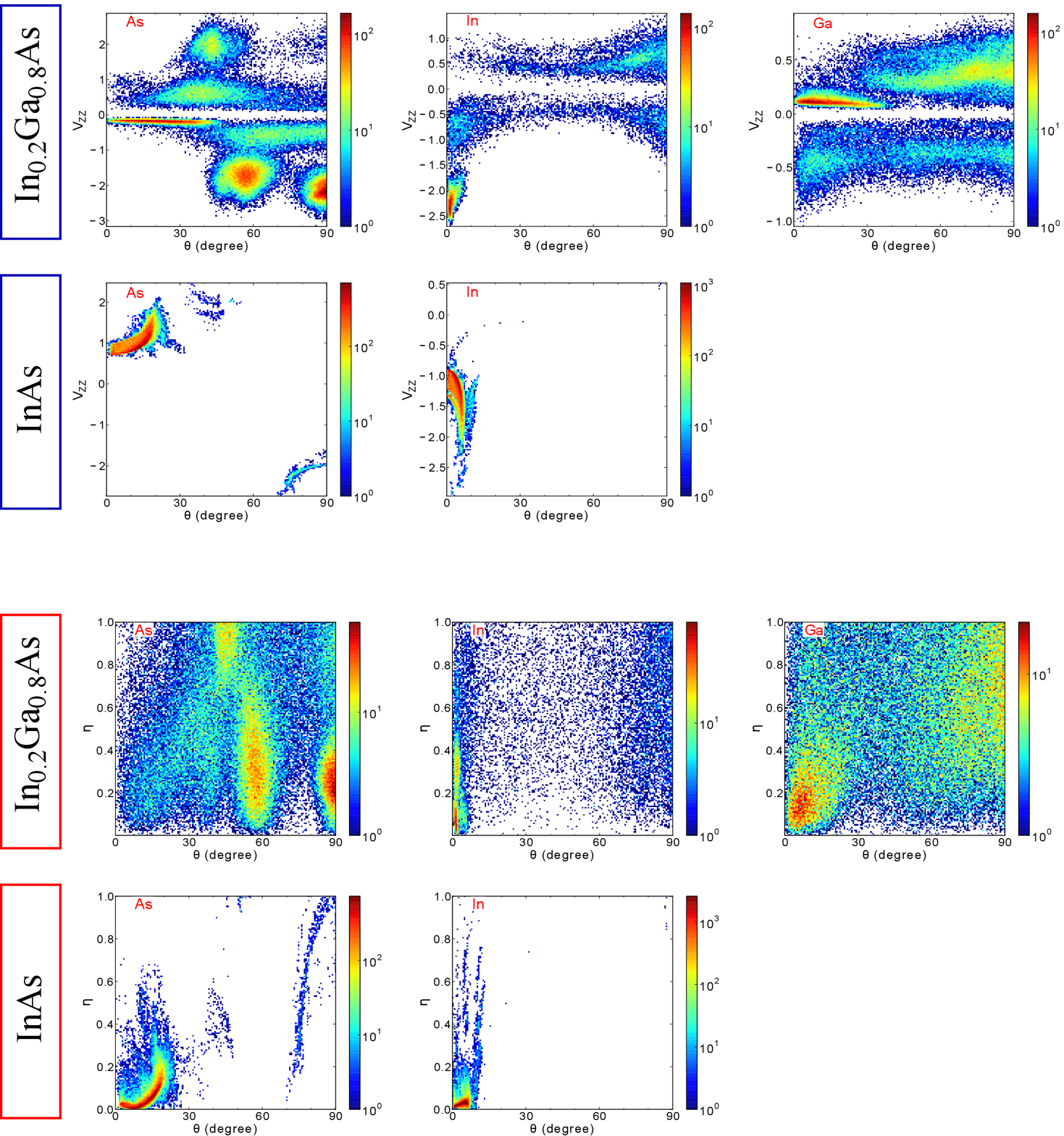}
\caption{(Color online) Two-dimensional EFG histograms. (Top two rows) EFG axial tilting, $\theta$ 
versus the major EFG, $V_{ZZ}$ (in units of $3\times 10^{21}$~V/m$^2$).
(Bottom two rows) EFG axial tilting, $\theta$ versus the EFG biaxiality, $\eta$. 
In each group, (upper rows) In$_{0.2}$Ga$_{0.8}$As QD; (lower rows) binary InAs QD. 
Color code represents the number of nuclei in the logarithmic scale.}
\label{histogram-2d}
\end{figure*}

\begin{figure*}
 \includegraphics[width=0.84\paperwidth]{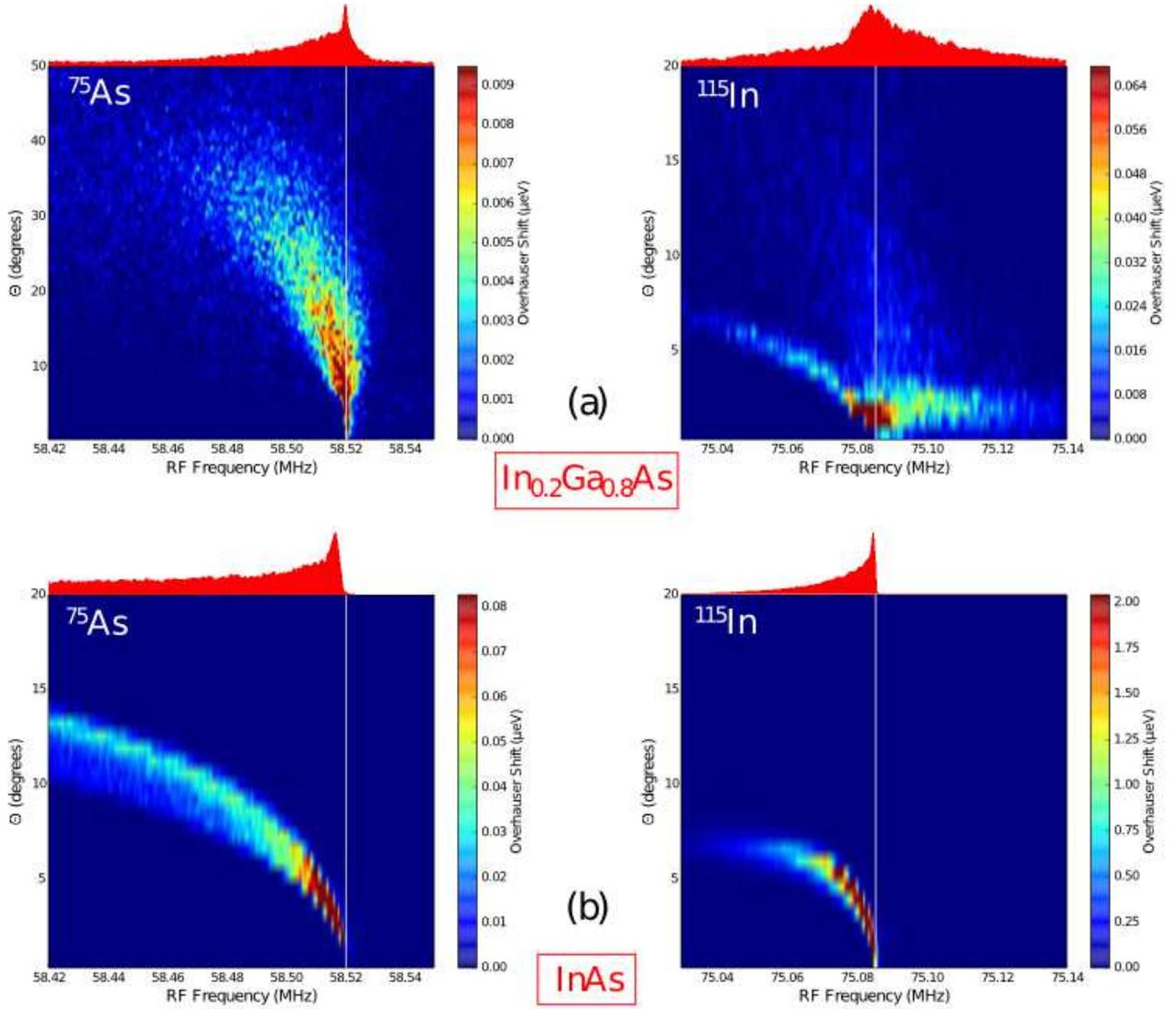}
\caption{(Color online) Same as Fig.~\ref{EFG-2d}, but with $B_0=8$~T, $f_{\hbox{\tiny gap}}$=1~kHz
to focus on CT. The color scale maxima are set to a quarter of their ordinary values to improve visibility.
For information purposes, full isotopic spectra, filled in red are placed on top; the vertical thin white 
line indicates the position of the pure Zeeman frequency for each case.
}
\label{EFG-2d-CT}
\end{figure*}

\begin{figure}
 \includegraphics[width=\columnwidth]{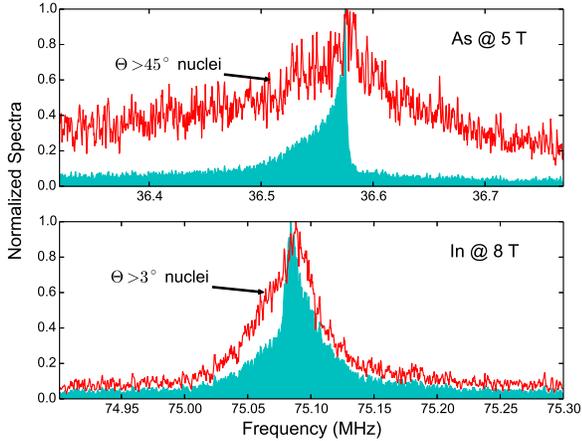}
\caption{(Color online) The effect of selected nuclei on the CT asymmetry based on their 
EFG axial tilting away from the static magnetic field, for the As nuclei at 5~T (top), and 
In nuclei at 8~T. Painted curves show the original contribution with all the nuclei.
For each case maxima are normalized to unity to assist the asymmetry comparison.
Alloy In$_{0.2}$Ga$_{0.8}$As QD is considered.
}
\label{excluded-nuclei-asymmetry}
\end{figure}

\begin{figure}
 \includegraphics[width=0.95\columnwidth]{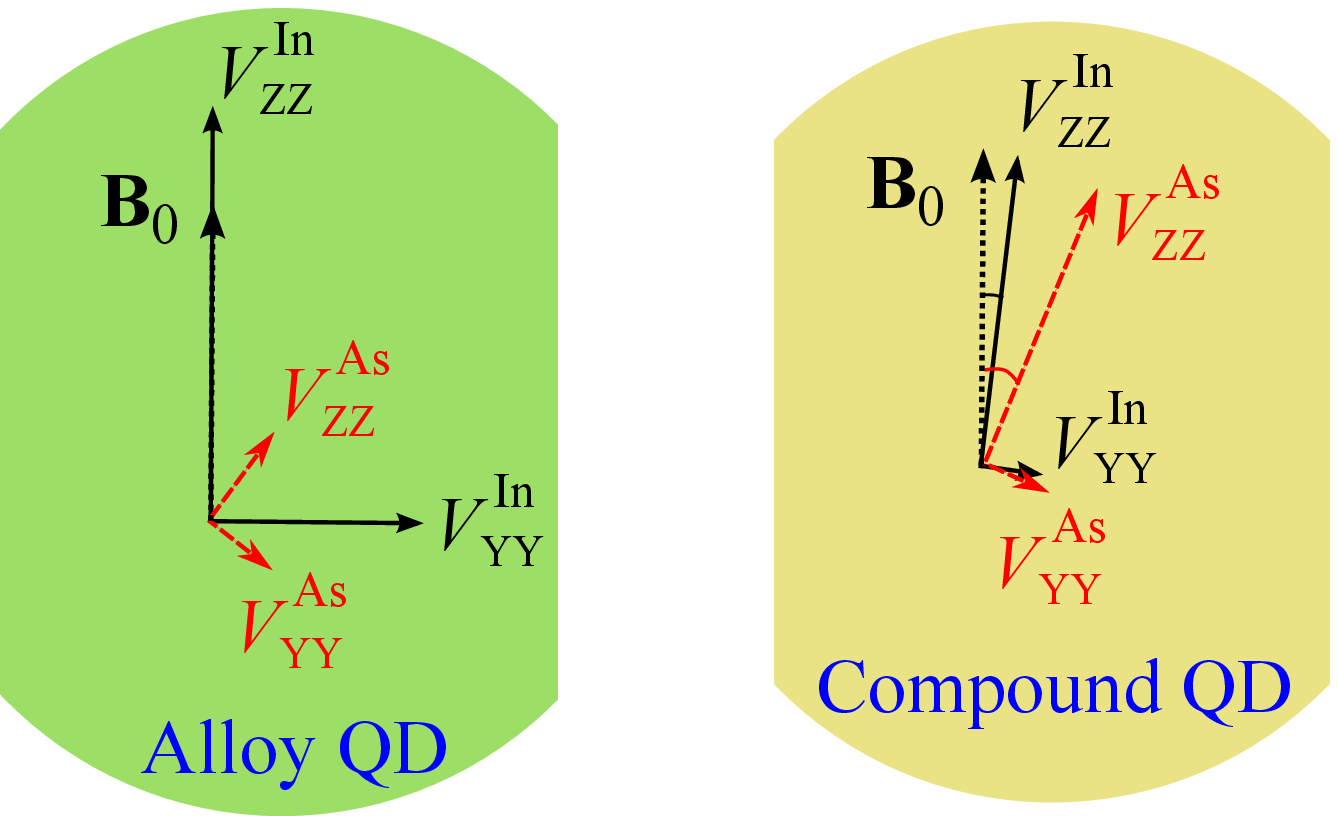}
\caption{(Color online) A schematic illustration for the typical EFG components and orientations 
for the indium (black/solid arrows) and arsenic (red/dashed arrows) nuclei of alloy In$_{x}$Ga$_{1-x}$As vs 
compound InAs QDs.
}
\label{EFGsummary-alloy-binary}
\end{figure}

\subsection{Opposite skewness of As and In CT lineshapes for the random alloy QD}
Based on the foregoing analysis, we can now address the intriguing
contrast between the CT lineshape asymmetries of the In and As nuclei as displayed 
in Figs.~\ref{CT-isotopes} and \ref{moments}. That is, the CT profile of In has positive skew 
(blue-tailed), while that for As has negative skew (red-tailed). 
The origin of these opposite behaviors is rooted in
the corresponding disparity in their EFG characteristics manifested by two-dimensional 
histograms in Fig.~\ref{histogram-2d}. Specifically, for the case of As, the group of nuclei strongly 
tilted with $\theta\lesssim45^\circ$ and $V_{ZZ}\lesssim 0$, and for the case of In those 
almost untilted $\theta\simeq 0^\circ$ but with large $|V_{ZZ}|$ are responsible for the opposite skewness. 
For the latter, this large $|V_{ZZ}|$ occurs from those In atoms residing in a relatively large 
biaxial strain environment, unlike the As or Ga atoms in relation to their CT. To reconcile these
features with the displayed skewness patterns we make use of a single-nucleus analysis. 
Refering to Appendix~D, EFG axial tilt at a constant $|V_{ZZ}|$ causes a red shift of the CT as 
the tilt increases up to $45^\circ$; on the other hand for the case of nearly untilted nuclei, 
increasing $|V_{ZZ}|$ and/or increasing EFG biaxiality, $\eta$ both cause a blue shift 
of the CT. 

To further substantiate these points, in Fig.~\ref{EFG-2d-CT}
we present major EFG orientation-resolved inverse spectra analysis focusing on the CT.
Here, the top left panel shows that the red-tailed skewness in As results from the tilted nuclei of 
an angle less than $45^\circ$, while on the top right panel the blue-tailed skewness of In 
is caused by almost untilted nuclei, $\theta<3^\circ$ (mind the different scales for the vertical axes). 
Note that the tilted In nuclei still give rise to red skew but they do not dominate.
Indeed, in Fig.~\ref{excluded-nuclei-asymmetry} we validate that if As (In) 
nuclei with tilt angle greater than $45^\circ$ ($3^\circ$) are only considered, the red 
(blue) skewness disappears. Next, returning to the bottom row of Fig.~\ref{EFG-2d-CT}
in the case of the binary InAs QD both As and In CT are red skewed, which can also be 
inferred from their similar two-dimensional EFG histograms in Fig.~\ref{histogram-2d} especially noting
the fact that both ensembles have highly uniaxial EFG. 
Again invoking the single-nucleus insight from Appendix~D, an increase in either the angular 
tilt or the $|V_{ZZ}|$ both work in the same direction leading to red shift of the CT.

\begin{figure*}
\hspace*{-1.6cm}
 \includegraphics[width=0.84\paperwidth]{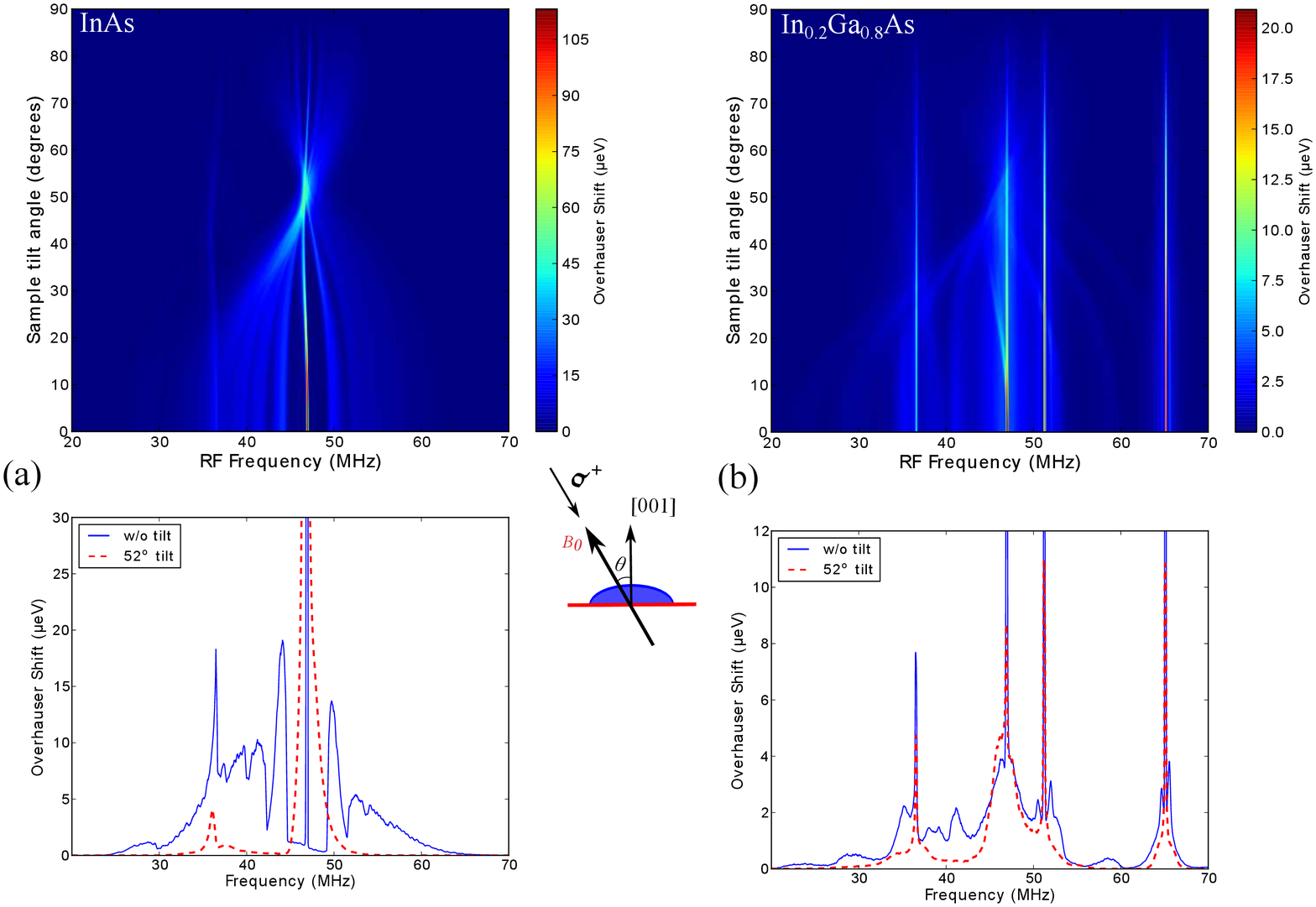}
\caption{(Color online) The effect of sample tilting on the inverse spectra.
(a) Binary InAs, (b) alloy In$_{0.2}$Ga$_{0.8}$As QD, both for
$B_0=5$~T, $f_{\hbox{\tiny gap}}$=200~kHz, $T_{\hbox{\tiny nuc}}$= 3~mK.
Faraday geometry is preserved, i.e., optical pumping and magnetic field are parallel, 
whereas the sample (growth axis) is tilted away from these, as shown in the inset.
Upper row displays the two-dimensional spectra of the tilting angle versus RF frequency;
lower row shows the one-dimensional spectra at zero and 52$^\circ$ tilting angles.
}
\label{2d-tilting-scan}
\end{figure*}

The typical EFG configurations responsible for the In and As CT lineshapes are summarized in 
Fig.~\ref{EFGsummary-alloy-binary}. In the alloy QD, strain inflates the EFG biaxiality
of both indium and arsenic nuclei; at the same time, in the case of arsenic enhances tilting,
and in the case of indium boosts the major EFG. The contrast between these two nuclei disappears 
in the compound QD. Both have quite uniaxial EFG with some axial tiltings:
distribution's mode for In is $\sim 5^\circ$ and for As it is $\sim 16^\circ$.

\subsection{Satellite transition collapse under sample tilting}
Other two-dimensional spectra can be generated with 
the added degree of freedom being the tilting of the sample growth axis with respect to 
the static magnetic field which is also taken to be the direction of the optical 
pump beam. The resultant spectra for binary InAs, and alloy In$_{0.2}$Ga$_{0.8}$As QD 
are shown in Fig.~\ref{2d-tilting-scan}, where we observe that around a tilting 
range of $50^\circ - 54^\circ$, the STs ``collapse'' 
on to CT, i.e., rendering all neighboring transitions energetically almost identical. 
Not surprisingly this is more distinctly the case for the compound InAs QD. 
To explain this behavior, in Appendix~D we demonstrate on a single $^{115}$In nucleus
how the opposite ST shifts disappear at a tilting angle ranging
between $45^\circ$ and $54.5^\circ$ depending on the biaxiality, $\eta$ 
(cf. Fig.~\ref{single-nucleus}). 
One outcome of this collapse is the reduction of the energy mismatch among the 
nuclei of the same isotope. Hence, especially for the indium 
nuclei which have the largest spin-9/2 manifold giving rise to the widest spread in energy, 
one can expect a shorter $T_2$ due to enhanced spin diffusion
under the critically-tilted angle ($\sim 52^\circ$) with respect to no-tilted case.
In a sense, sample tilting together with dipole-dipole interaction can act like a 
\textit{flush mechanism} for evening out population imbalances among nuclear spin states
brought and sustained by optical orientation.

\section{Conclusions}
This computational study demonstrates the power of the NMR inverse spectra 
technique as a tool for retrieving atomistic level structural information from 
strained SAQDs. Through a comparative assessment of alloy versus compound In(Ga)As QDs,
as well as of the involved nuclear species, we unveil marked differences in 
their spectral features, and establish links with the local chemical 
structure, strain, and material properties. Our main findings are grouped as follows:
(i)~\textit{Strain and CT asymmetry};
In compound InAs QDs the dominant component is the compressive biaxial strain which causes quite a 
uniaxial and rather strong EFG. The shear strain has a secondary role being significant 
around the interfaces and results in a limited, yet still crucial EFG axial tilting. Indium and 
arsenic nuclei qualitatively both obey this picture. The situation becomes more complex
with the sway of random alloying in the InGaAs QD. The shear strain spreads all throughout 
the core and plays a primary role. Moreover, EFG biaxiality of all elements get enhanced. 
An interesting aftermath of atomistic alloy strain is that
the indium nuclei are mostly untilted but have large EFG values, whereas arsenic
nuclei have low EFG values but with excessive axial tiltings. Under
the realm of these different EFG conditions, the asymmetry of 
the CT in compound QDs gets red-skewed for both In and As, while for the alloy QD
the In CT acquires a dominant blue tail arising from a large untilted In nuclear 
population residing in a relatively large biaxial strain environment.
(ii)~\textit{Arsenic and cation alloying};
Compared to In and Ga nuclei, As bares a number of distinctions.
First, due to its low gyromagnetic ratio As nuclei are more prone to QI
under a given magnetic field compared to In or Ga. Secondly, the shear strain is 
most operative on the arsenic nuclei. This stems from the large $S_{44}$ component of 
the gradient elastic tensor of arsenic nucleus, as well as nearest-neighbor variations 
because of cation alloying, not present for In and Ga categorically. 
An implication of these is that if there exists an alloy structure within the QD region,
this can be identified, in principle by an additional peak in the arsenic ST. Specifically, it 
originates from those arsenic nuclei with their major quadrupolar EFG axes tilted 
perpendicular to the growth axis, a direct outcome of alloying.
(iii)~\textit{Sample tilt};
Finally, we predict the collapse of the STs onto CT which is most pronounced in compound InAs QDs, 
and the possibility of negating QI and restoring a monoenergetic distribution like a solitary 
Zeeman interaction, simultaneously for all isotopic nuclear spins by tilting the sample about 
52$^\circ$ with respect to static magnetic field. 
These findings must be verified experimentally and superseded by further studies for the
purposes of both atomistic material insights and also for the coherent control of a relatively 
small number of nuclear spins embedded in a strained confined environment.

\begin{acknowledgments}
C.B. would like to thank T\"UB\.ITAK, T\"{u}rkiye Bilimsel ve Teknolojik Ara\c{s}t{\i}rma 
Kurumu, for financial support through Project No. 112T148.
E.A.C. was supported by a University of Sheffield Vice-Chancellor's Fellowship.
All authors acknowledge the support of UK Royal Society International Exchanges.
A.I.T. and E.A.C. thank the financial support of the EPSRC Programme Grant EP/J007544/1.
\end{acknowledgments}

\appendix
\section*{Appendix~A: Nuclear polarization along the static magnetic field}
\begin{figure}
 \includegraphics[width=1.05\columnwidth]{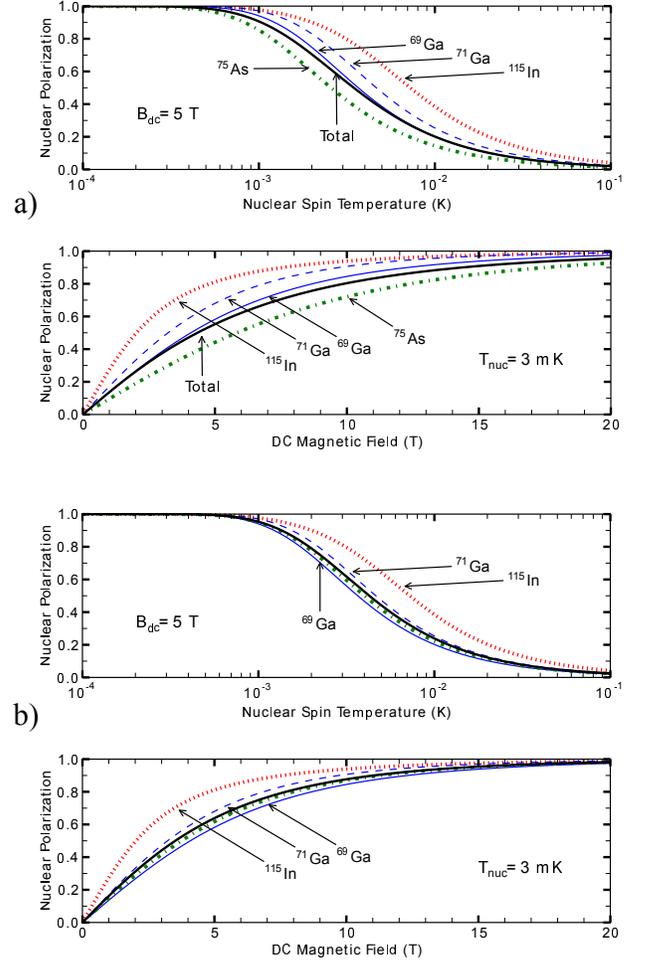}
\caption{(Color online) The isotope-resolved and total nuclear spin polarizations along 
the static magnetic field for the alloy In$_{0.2}$Ga$_{0.8}$As QD. (a) Using the original
gyromagnetic ratio for arsenic, (b) with the $^{75}$As gyromagnetic ratio artificially
increased to the average value of $^{69}$Ga and $^{71}$Ga, 
i.e., $\gamma_{As}\rightarrow\gamma_{Ga}$. Linestyle sets are the same for each panel.}
\label{polarization}
\end{figure}

In this appendix, we discuss the average nuclear spin orientation for each  
isotopic ensemble within the alloy In$_{0.2}$Ga$_{0.8}$As QD in the presence of both 
Zeeman and QI terms. Even though the results here are qualitatively
along the normal expectations, the quantitative details and isotopic variations 
may still be worthwhile considering.
In Fig.~\ref{polarization} we show the average nuclear polarization along the static magnetic field, 
$P^e$ for each isotopic ensemble as well as the total values, based on the weighted contribution
of each isotope within the QD; 
in part (a), the upper plot depicts nuclear 
spin temperature dependence at a fixed magnetic field of 5~T, whereas the lower plot shows 
how it varies with respect to magnetic field at a fixed nuclear spin temperature, 
$T_{\hbox{\tiny nuc}}=3$~mK. The latter depends on the optical pumping and nuclear spin energetics.
Qualitatively, in both cases all isotopes display the expected polarization trends under increased 
magnetic field or decreased nuclear spin temperature.
On the quantitative side, there is a striking difference among the elements, 
namely arsenic nuclear ensemble's polarization is substantially lower than the other elements.
This is the origin of the mentioned smallest CT peak of arsenic among all elements 
in Fig.~\ref{isotope-helicity}.
Stated quantitatively, at $B_0=5$~T and $T_{\hbox{\tiny nuc}}=3$~mK, arsenic 
nuclear spin polarization is about 44\%, whereas $^{69}$Ga $\rightarrow$ 58\%, 
$^{71}$Ga $\rightarrow$ 68\%, and $^{115}$In $\rightarrow$ 81\%. With the largest 
population belonging to the arsenic nuclei due to alloy partitioning between cations, 
the overall average polarization value (denoted as total in Fig.~\ref{polarization}) 
comes out as 55\%, i.e., closer to that of the arsenic value.

\begin{figure}
 \includegraphics[width=0.95\columnwidth]{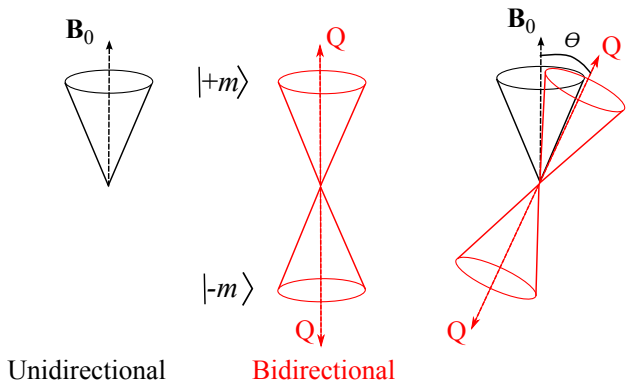}
\caption{(Color online) Zeeman interaction having a unidirectional axis (left) versus QI 
with a bidirectional character (center). In actual QD samples the local major 
quadrupole axis is somewhat tilted because of shear strain with respect to magnetic 
field in Faraday geometry (right), which can accordingly reduce their competition.}
\label{uni-bi} 
\end{figure}

\textit{Why is it harder to orient the arsenic nuclei?}
By far the most critical factor is the gyromagnetic ratio, $\gamma$ 
which is substantially smaller for arsenic compared to other elements. 
To illustrate this point, we artificially increase  $\gamma$ for $^{75}$As 
by 55\% so that it reaches to the average value of the $^{69}$Ga and 
$^{71}$Ga ensemble. It can be observed in part (b) of Fig.~\ref{polarization} 
that arsenic as well as the total nuclear polarization now lie 
in between the $^{69}$Ga and $^{71}$Ga curves.
In particular, for $B_0=5$~T and $T_{\hbox{\tiny nuc}}=3$~mK, 
the arsenic spin alignment increases from its actual value of 44\% to 62\%.
The underlying reason is that QI has a {\em bidirectional} character; being an 
electrostatic interaction in nature, it cannot discriminate the states $|\pm m\rangle$,
whereas Zeeman interaction being {\em unidirectional} splits the $|+m\rangle$ and $|-m\rangle$ 
states thereby promoting nuclear spin polarization (see, Fig.~\ref{uni-bi}).
In other words, while the static magnetic field (Zeeman term) tries to polarize the 
nuclear spins, QI tries to erase this.
Hence, with their low $\gamma$ value, $^{75}$As nuclei are more prone to the quadrupolar 
\textit{depolarization} compared to other isotopes at the same external magnetic field. 
In the case for pure InAs QD (not shown), the trends are similar, however, 
as this QD is much more strained, here the relevant component is the biaxial strain, 
$\epsilon_B$,\cite{bulutay12} the quadrupolar effects are somewhat more pronounced.

\begin{figure*}
 \includegraphics[width=0.75\paperwidth]{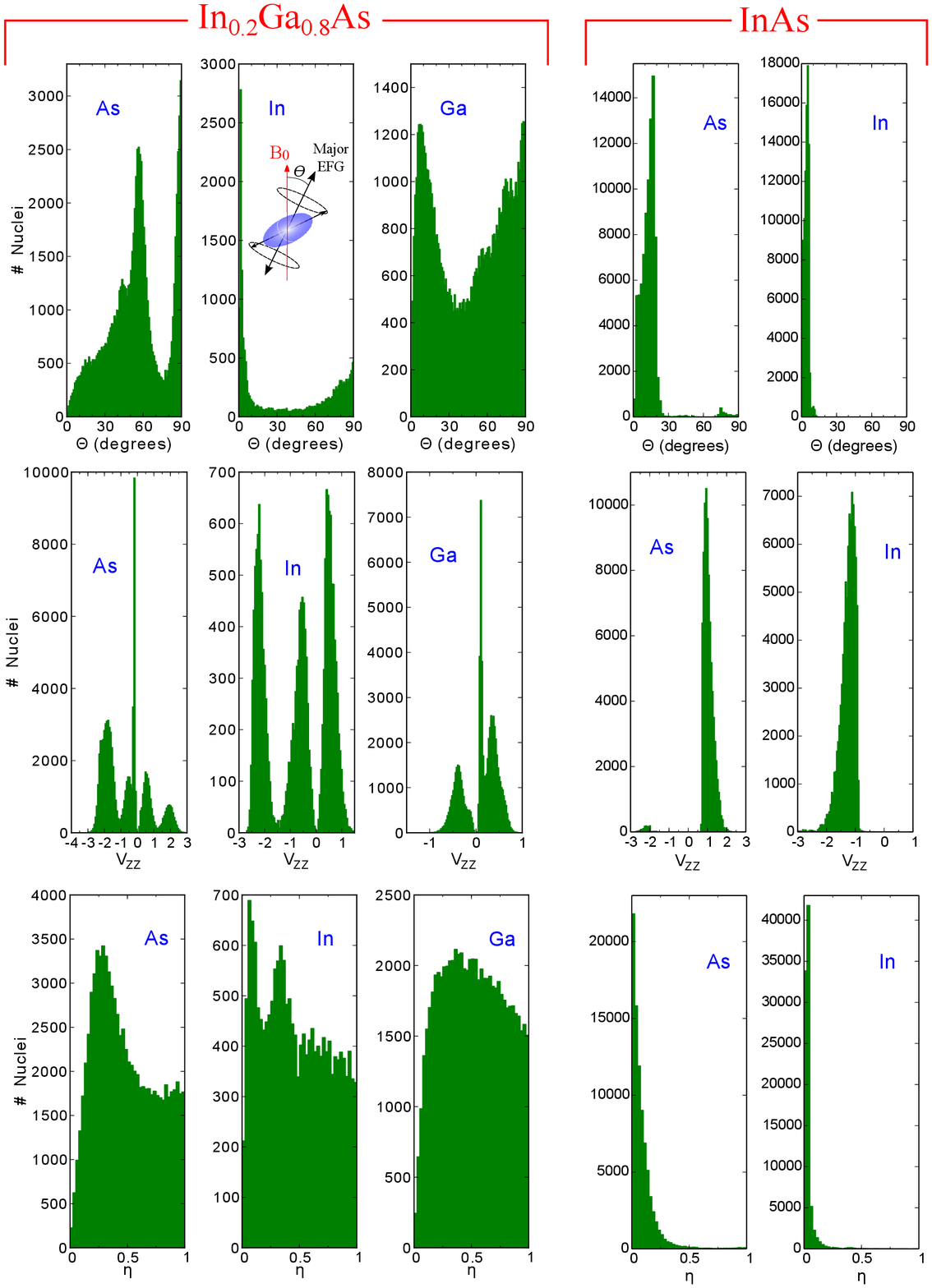}
\caption{(Color online) For the alloy In$_{0.2}$Ga$_{0.8}$As and binary InAs QD histograms 
of the three EFG parameters. (Top row) The angular tilting of the major EFG axis 
away from the magnetic field, $\theta$ (inset). (Middle row) Major EFG, $V_{ZZ}$ in 
units of $3\times 10^{21}$~V/m$^2$. (Bottom row) EFG biaxiality, $\eta$.
The results here refer to the case after the QD strain relaxation.}
\label{histogram}
\end{figure*}

\section*{Appendix~B: Histograms for the alloy and the compound QD}
\begin{figure}
 \includegraphics[width=0.75\columnwidth]{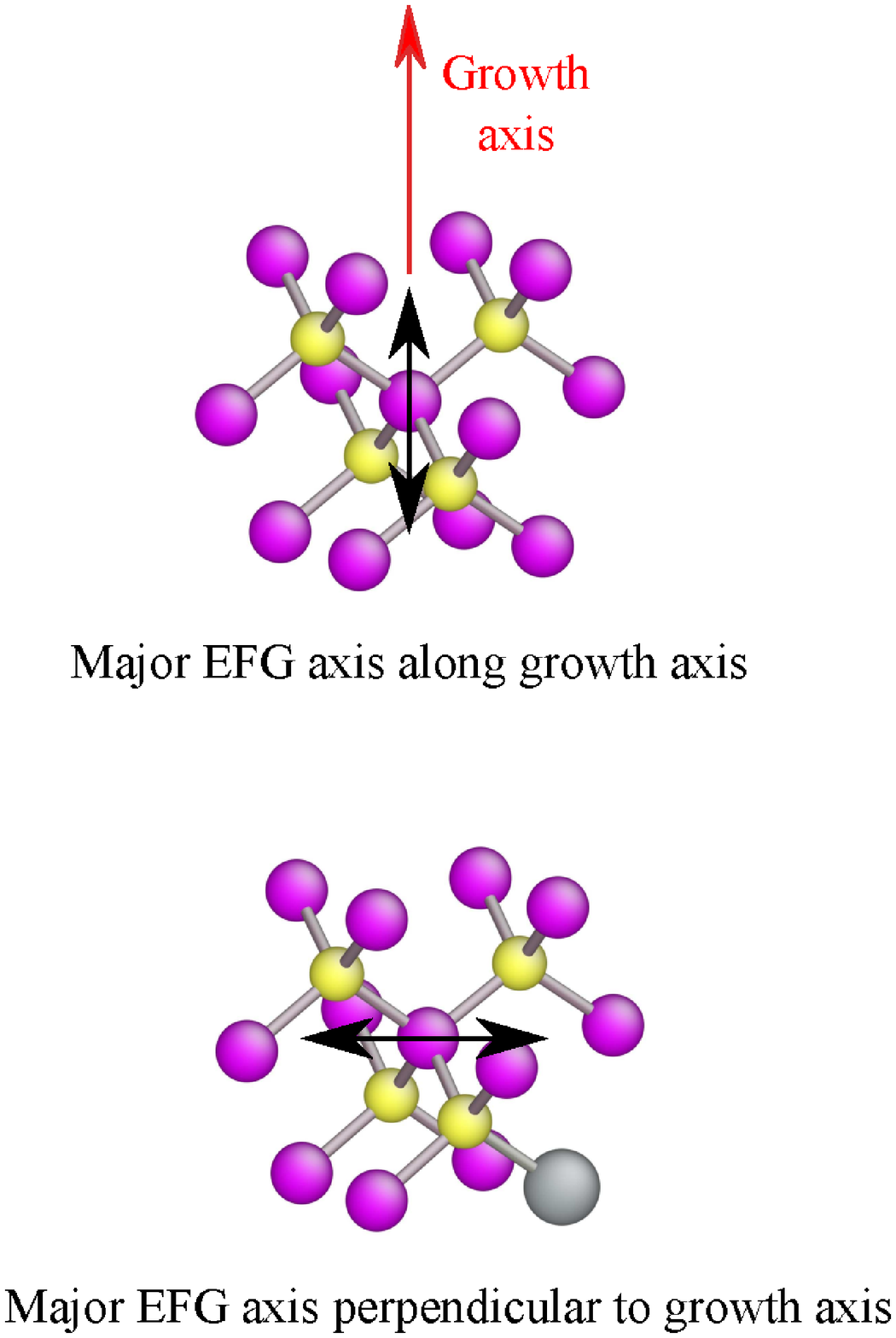}
\caption{(Color online) The switching of the major EFG axis of the center gallium nuclei 
(indicated by black double arrows) from parallel (top)
to perpendicular (bottom) orientation with respect to the growth axis (also the direction 
of the static magnetic field) by a change in a second-nearest-neighbor atom. Color coding 
is as follows: indium in gray, gallium in purple, arsenic in yellow.}
\label{alloy-configs}
\end{figure}

In this appendix, histograms for the three EFG parameters, namely,
major quadrupole axial tilting away from the static magnetic field, $\theta$,
the value of the major EFG, $V_{ZZ}$, and the EFG biaxiality, $\eta$ are discussed
comparatively for the alloy and compound QDs. Starting with axial tilting for the alloy 
In$_{0.2}$Ga$_{0.8}$As QD as displayed in the top row of Fig.~\ref{histogram}, we observe 
a remarkable dissimilarity in the orientations of arsenic nuclei 
in comparison to cations (Ga, In). The latter display somewhat similar characteristics, 
that peaks either along or perpendicular to the magnetic field. As mentioned in the main text, 
for the cation nuclei (Ga, In) the quadrupole axial tilting is driven
by a change in their next-nearest neighborhoods. All together there are 144 configurations;
we show in Fig.~\ref{alloy-configs} two such gallium-centric instances where a change in the 
second-nearest neighbor atom switches the major EFG axis from parallel to perpendicular orientation
with respect to growth axis. 
The distribution of the axial tilting of As nuclei has two peaks at 
45$^\circ$ and 54.7$^\circ$ that coincide with those pre-relaxed configurations of 
Table~\ref{arsenic-configs}, even obeying the same 1:2 ratio of the relative weights of 
these cation-bonding orientations. Next, considering the binary InAs QD, due to lack 
of alloying for this case, both In and As nuclei's major EFG axes are more or less 
aligned along the magnetic field. The interface As atoms which are relatively low in 
number do still have mixed cation neighbors and this gives rise to some limited 
variance and axial tilting. Analyzing the middle rows,
in the alloy QD major EFG values are evenly distributed on either side of zero, 
and grouped in a few bunches; for the compound QD, the nuclei are gathered around a single
$V_{ZZ}$ value with opposite signs for In and As. The biaxiality parameters $\eta$ (bottom rows)
of the two QDs are also markedly different: They are spread over the full accessible 
range for the alloy QD, whereas the compound QD EFG is quite uniaxial, mainly restricted 
to lower than the 0.25 value.

\section*{Appendix~C: The role of quadrupole parameters on the lineshape} 
\begin{figure}
 \includegraphics[width=1.05\columnwidth]{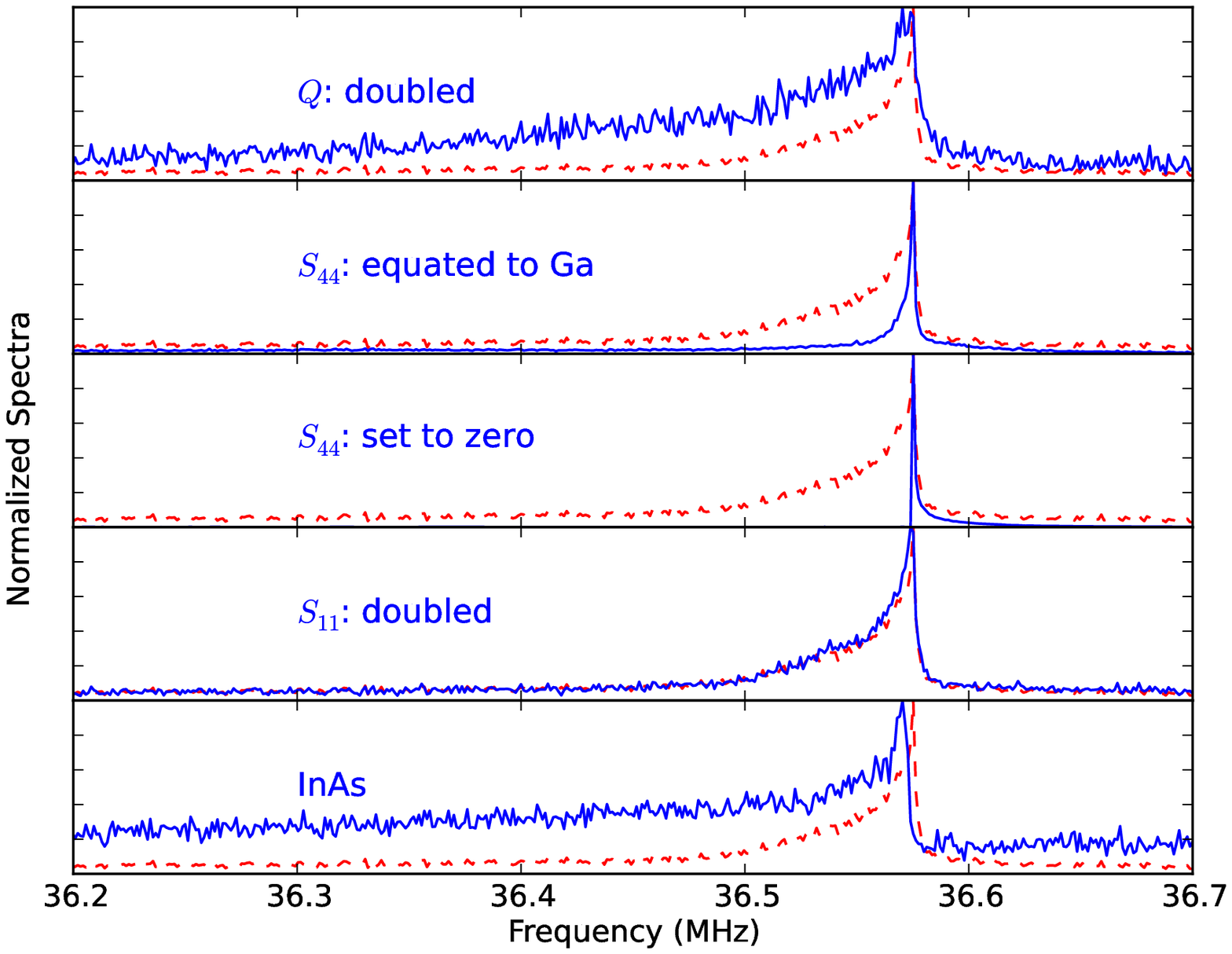}
\caption{(Color online) Dependence of arsenic resonance of CT asymmetry on the nuclear quadrupole 
parameters and alloy mole fraction. In all panels dotted (red) lines refer to original 
In$_{0.2}$Ga$_{0.8}$As QD with 
$f_{\hbox{\tiny gap}}$=1~kHz, $B_0=5$~T, $T_{\hbox{\tiny nuc}}$= 3~mK, 
$\sigma^{+}$ optical pumping. Solid (blue) lines demonstrate the cases after a modification in material 
parameters, $Q, S_{44}, S_{11}$, as well as
the case for binary InAs QD.  Each peak is set unity to compare the lineshapes.}
\label{CT-material}
\end{figure}

The aim of this appendix is to develop a feeling for the importance of the individual 
nuclear quadrupole parameters ($Q, S_{11}, S_{44}$) of As in regard to CT lineshape.
As illustrated in Fig.~\ref{CT-material}, if we double the electric quadrupole moment $Q$, 
the red-tailed asymmetry is enhanced indicating that its origin is the QI. In particular, the 
$S_{44}$ component of the gradient elastic tensor primarily controls the asymmetric profile:
Lowering this value to that of Ga (i.e., decreasing by about 
2.5 times) drastically reduces the asymmetry, while setting it to zero totally removes 
and even reverses its direction. Since $S_{44}$ relates the off-diagonal entries of EFG and the
strain tensors, its effectiveness directly invokes to the importance of shear strain on the CT asymmetry.
On the other hand, the $S_{11}$ component is not functional, doubling its value virtually leaves 
the asymmetry unchanged. 

\begin{figure}
 \includegraphics[width=0.95\columnwidth]{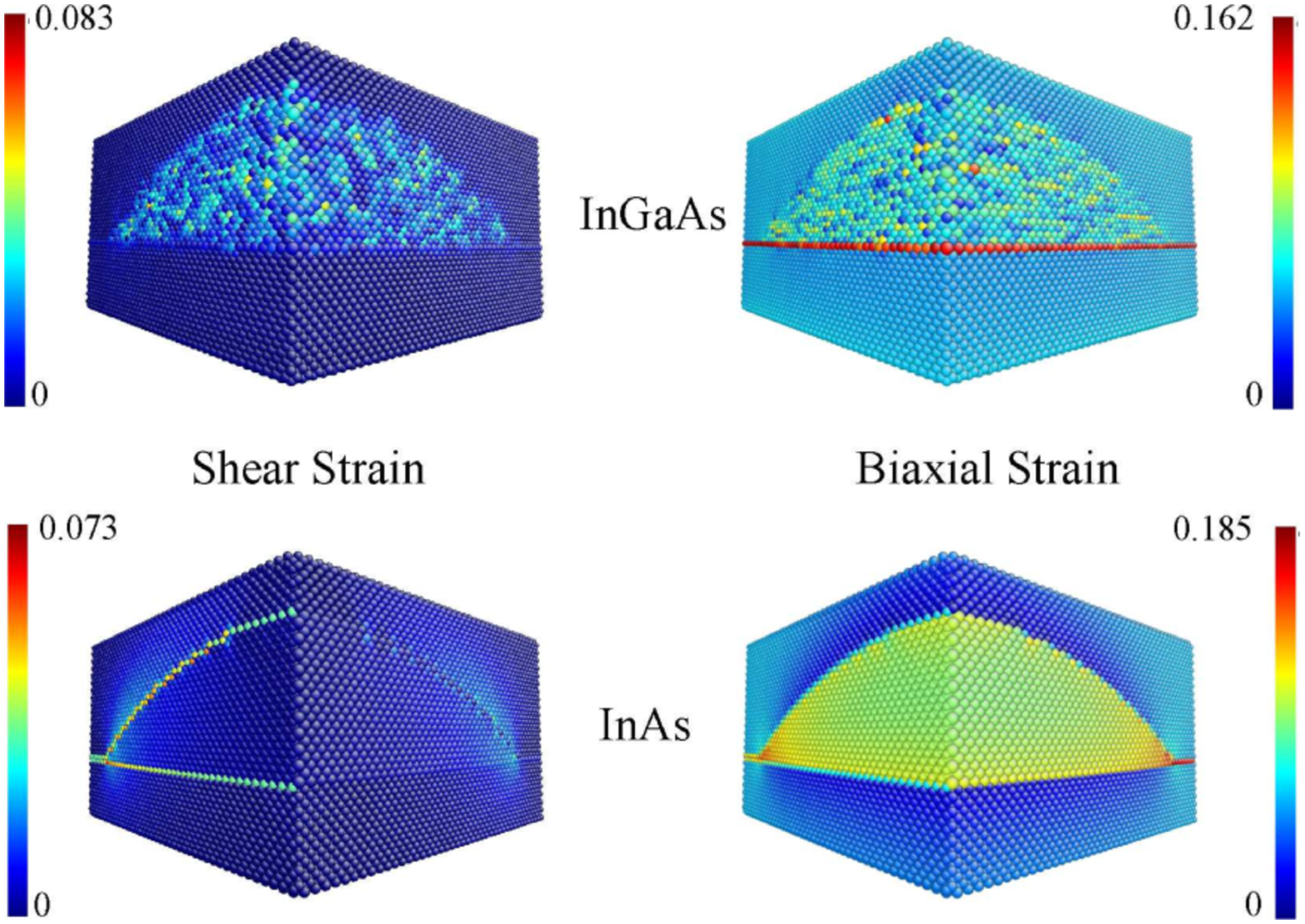}
\caption{(Color online) Atomistic shear, $\epsilon_S$ (left) and biaxial, $\epsilon_B$ (right) 
strain distributions for the In$_{0.2}$Ga$_{0.8}$As (top) and InAs (bottom) QDs, cut through 
both (100) and (010) planes. The lens-shaped QD boundaries can easily be identified on the 
InAs QD from the enclosing bright shear strain regions corresponding to the interfaces
with the host matrix as well as the wetting layer.}
\label{strain-profiles} 
\end{figure}

Note that all of the above statements refer to the In$_{0.2}$Ga$_{0.8}$As QD.
Additionally, in the bottom panel of Fig.~\ref{CT-material} we compare the alloy QD 
with the binary InAs QD, where in the latter the CT asymmetry of As nuclei gets significantly enhanced. 
This is at odds with the established insight so far based on the prime 
importance of the shear strain, in conjunction with its small value for binary InAs QD. As a matter of
fact as shown in the left column of Fig.~\ref{strain-profiles}, the interior of the 
In$_{0.2}$Ga$_{0.8}$As QD retains an exuberant shear strain profile due to random alloying, whereas
this gets diminished in the core of the InAs QD and only becomes significant toward the interfaces. 
However, a change 
of roles is observed in the right column of the same figure where the compound QD has much 
stronger biaxial strain compared to alloy QD simply due to larger lattice mismatch between the 
core and the matrix regions. As we have stated in the atomistic analysis (see, also Appendix~D), 
not only shear but also the biaxial strain component can cause a shift in CT; the former acts
through EFG axial tilting as in alloy QD, and the latter via the major EFG value in the case of
compound QD.

\section*{Appendix~D: Single-nucleus particular EFG parameter traits}
In this appendix considering a single nucleus governed by the general Hamiltonian, 
$\oH_Q + \oH_M$, we present how CT and ST frequencies shift under various 
combinations of the three EFG parameters: 
the major EFG value ($V_{ZZ}$), the angular deviation of the major EFG 
axis from the static magnetic field ($\theta$), and biaxiality ($\eta$). 
As the trends are qualitatively similar among
the nuclear species of this work, for demonstration purposes we choose $^{115}$In. 
The static magnetic field is taken as 8~T which is used in some of our calculations 
in the text. Within the STs, we consider the $3/2 \leftrightarrow 1/2$ transition. 

\begin{figure}
 \includegraphics[width=1.05\columnwidth]{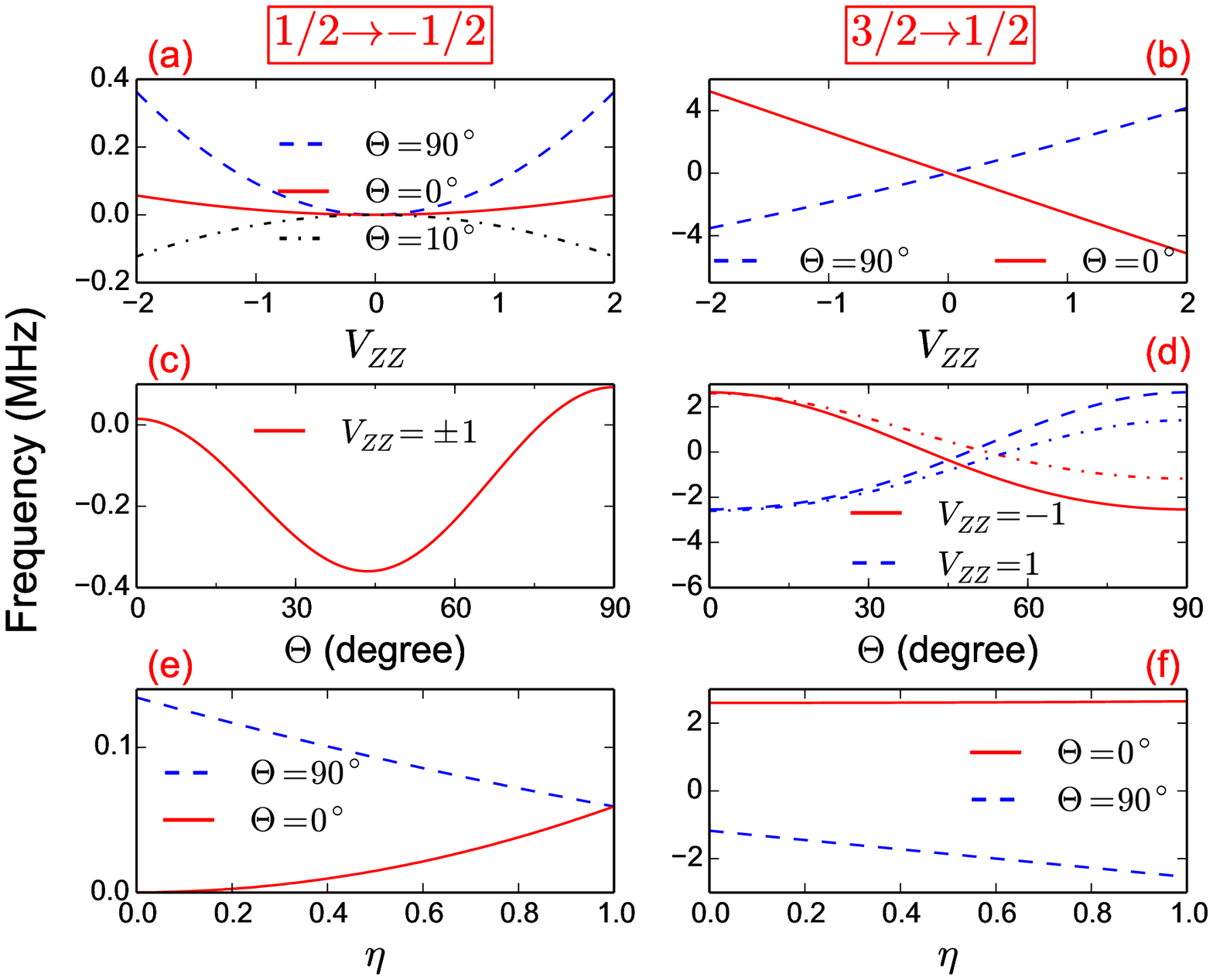}
\caption{(Color online) Demonstration on a single $^{115}$In nucleus of the effects of
various EFG parameter combinations on the CT (left) and ST (right) frequencies.
For convenience Zeeman frequency is subtracted to highlight QI.
The major EFG, $V_{ZZ}$ is in units of $3\times 10^{21}$~V/m$^2$.
(a)--(c) are for $\eta=0.5$; (d) shows both $\eta=0$ and $\eta=1$ 
cases with the latter in dash-dotted lines; (d) and (e) are for $V_{ZZ}=-1$.
}
\label{single-nucleus} 
\end{figure}
Figure~\ref{single-nucleus} displays the variation with respect to each one 
of $V_{ZZ}$, $\theta$, $\eta$ while keeping the other two parameters fixed.
As the shift of CT is in second order under QI,\cite{cohen57} the dependence 
on $V_{ZZ}$ is quadratic, hence independent of its sign [Fig.~\ref{single-nucleus}(a)]; 
in contrast, ST shifts being first-order are much stronger [Fig.~\ref{single-nucleus}(b)];.
For both CT and ST the direction of shift depends on $\theta$.
In the case of CT for small angles there is a blue shift with increasing $|V_{ZZ}|$,
which becomes a red shift for larger angles, Fig.~\ref{single-nucleus}(a);, 10$^\circ$ curve. 
For the specific case considered here ($^{115}$In and $\eta=0.5$) this transition 
occurs at $\theta =5.55^\circ$. Yet, for even larger angles this reverts back 
to a blue shift, Fig.~\ref{single-nucleus}(a), 90$^\circ$ curve.
The continuous variation of $\theta$ produces a cosine-type shift in either CT
and ST, again with the effect being much stronger for the latter, 
Figs.~\ref{single-nucleus}(c) and (d). 
Observe that the shifts in ST for $\pm V_{ZZ}$  cross each other at a $\theta$ value ranging
between $45^\circ$ and $54.5^\circ$ as $\eta$ varies from 0 to 1. Other STs
(for the case of In) display a similar pattern. It is this behavior that is harnessed in 
the collapse of STs at a convenient sample tilting around 52$^\circ$.
The bare dependence on $\eta$ can be seen in Figs.~\ref{single-nucleus}(e) and (f); 
in particular it shows a blue shift in CT for $\theta=0$.
A rather inhomogeneous mixture of these single-nucleus traits as governed by the
atomistic strain field gives rise to a unique fingerprint of the QD NMR.


\begin{thebibliography}{99}
\bibitem{note-notation}An alternative notation in the literature is the use of lowercase
subscripts, but with a `$P$' superscript, i.e., $V^P_{zz}$, to designate the EFG principal axis.
As we would like to reserve the superscript for the element/isotope label when necessary, we 
prefer to use $V_{ZZ}$.
\end{thebibliography}
\end{document}